\documentclass{article}
\usepackage{spconf,amsmath,graphicx,hyperref}
\usepackage[nolist, nohyperlinks]{acronym}
\usepackage{nicefrac}
\usepackage{cleveref}
\usepackage{booktabs}
\usepackage{pifont}
\usepackage{cite}

\usepackage{circuitikz}
\usepackage[dvipsnames]{xcolor}
\usepackage{tikz}
\usetikzlibrary{decorations.pathreplacing,calligraphy, arrows.meta, shapes.geometric, positioning, shapes.misc, calc}
\usepackage{pgf}
\usepackage{pgfplots}
\usepackage{contour}
\contourlength{0.4pt}
\tikzset{cross/.style={cross out, draw=black, minimum size=2*(#1-\pgflinewidth), inner sep=0pt, outer sep=0pt},}
\usepackage{subcaption}

\usepackage{siunitx}

\pgfplotsset{compat=1.18}

\DeclareMathAlphabet\mathbfcal{OMS}{cmsy}{b}{n}

\acrodefplural{nns}[NNs]{neural networks}
\acrodefplural{irm}[IRMs]{ideal ratio masks}
\acrodefplural{ibm}[IBMs]{ideal binary masks}
\acrodefplural{doa}[DoAs]{directions of arrival}
\acrodefplural{rirs}[RIRs]{room impulse responses}

\begin{acronym}
\acro{stft}[STFT]{short-time Fourier transform}
\acro{rir}[RIR]{room impulse response}
\acro{istft}[iSTFT]{inverse short-time Fourier transform}
\acro{doa}[DoA]{direction of arrival}
\acro{irm}[IRM]{ideal ratio mask}
\acro{mvdr}[MVDR]{minimum-variance distortionless response}
\acro{nn}[NN]{neural network}
\acro{dnn}[DNN]{deep neural network}
\acro{gcc_phat}[GCC-PHAT]{generalized cross-correlation phase transform}
\acro{ssl}[SSL]{sound source localization}
\acro{srp_phat}[SRP-PHAT]{steered response power with phase transform}
\acro{mse}[MSE]{mean squared error}
\acro{mae}[MAE]{mean angular error}
\acro{sdr}[SDR]{signal-to-distortion ratio}
\acro{sisdr}[SI-SDR]{scale-invariant signal-to-distortion ratio}
\acro{pesq}[PESQ]{perceptual evaluation of speech quality}
\acro{polqa}[POLQA]{perceptual objective listening quality}
\acro{wer}[WER]{word error rate}
\acro{pit}[PIT]{permutation-invariant training}
\acro{lbt}[LBT]{location-based training}
\acro{asr}[ASR]{automatic speech recognition}
\acro{bce}[BCE]{binary cross-entropy}
\acro{jnf}[JNF]{joint non-linear filter}
\acro{nn}[NN]{neural network}
\acro{dnn}[DNN]{deep neural network}
\acro{mccruse}[MC-CRUSE]{multi-channel convolutional recurrent U-net architecture for speech enhancement}
\acro{estoi}[ESTOI]{extended short-time objective intelligibility}
\acro{snr}[SNR]{signal-to-noise ratio}
\acro{sir}[SIR]{signal-to-interference ratio}
\acro{iid}[i.i.d.]{independent and identically distributed}
\acro{wrt}[w.r.t.]{with respect to}
\acro{ssf}[SSF]{spatially selective filter}
\acro{cv}[CV]{constant velocity}
\acro{rw}[RW]{random walk}
\acro{ar}[AR]{autoregressive}
\acro{tse}[TSE]{target speaker extraction}
\acro{tst}[TST]{target speaker tracking}
\acro{tsl}[TSL]{target speaker localization}
\acro{dnsmos}[DNSMOS]{deep noise suppression mean opinion score}
\acro{map}[MAP]{maximum a posteriori}
\acro{kf}[KF]{Kalman filter}
\acro{pf}[PF]{particle filter}
\acro{das}[DaS]{delay-and-sum}
\acro{ae}[AE]{angular error}
\acro{acc}[ACC]{accuracy}
\acro{miso}[MISO]{multiple-input and single-output}
\acro{mimo}[MIMO]{multiple-input and multiple-output}
\acro{foa}[FOA]{first order Ambisonics}
\acro{seld}[SELD]{sound event localization and detection}
\acro{mac}[MAC]{multiply-accumulate operation}
\acro{acn}[ACN]{Ambisonics channel number ordering}
\acro{rds}[RDS]{recurrent deep stacking}
\acro{mos}[MOS]{mean opinion score}
\acro{nisqa}[NISQA]{non-intrusive speech quality assessment}
\end{acronym}

\definecolor{tab_blue}{RGB}{31, 119, 180}
\definecolor{tab_green}{RGB}{44, 160, 44}
\definecolor{tab_orange}{RGB}{255, 127, 14}
\definecolor{tab_purple}{RGB}{148, 103, 189}
\definecolor{tab_brown}{RGB}{165, 42, 38}
\definecolor{tab_pink}{RGB}{227, 119, 194}
\definecolor{tab_cyan}{RGB}{23, 190, 207}
\definecolor{orig_green}{RGB}{44, 160, 44}
\definecolor{orig_purple}{RGB}{148, 103, 189}
\definecolor{custom_green}{RGB}{149.5, 207.5, 149.5}
\definecolor{custom_purple}{RGB}{201.5, 179, 222}
\definecolor{dark_gray}{RGB}{128, 128, 128}

\newcommand\tickWidth{0.5pt}
\newcommand\rectEdge{1mm}

\newcommand\elevation{\phi}
\newcommand\sphHarmonics{b}
\newcommand\sphBasis{B}
\newcommand\delayIdx[1]{
    \scriptstyle #1\hspace{0.25mm}\raisebox{0.4ex}{\rule{0.3em}{0.06ex}}1
}

\newcommand\projectPage{
    \scriptsize \texttt{https://sp-uhh.github.io/adaptive-rotary-steering/}
}

\newcommand\splitSize{3pt}
\newcommand\addSize{6pt}
\newcommand\shadowOff{0.375mm}
\newcommand\smallMath{\tiny}
\newcommand\myDelta{\Delta\hspace{-0.25mm}}
\newcommand\ambixWidth{0.5pt}
\newcommand\rotEllipseA{0.27cm} 
\newcommand\rotEllipseB{0.12cm}  
\newcommand{\rotSymb}[1]{
     \makebox[1cm][l]{\hspace{-1.15mm}\raisebox{0mm}{
        \begin{tikzpicture}
            \draw[x=\rotEllipseB, y=0.95*\rotEllipseA, line width=0.175ex, -stealth, color=lightgray] ({1*cos(-150)},{1*sin(-150)}) arc (-150:150:1 and 1);
            \draw[x=\rotEllipseA, y=0.7*\rotEllipseB, line width=0.175ex, -stealth, color=lightgray] ({1*cos(-150)},{1*sin(-150)}) arc (-150:150:1 and 1);
            \node[anchor=center] at (0,0) {\scriptsize\contour{white}{\hspace{0.5mm}#1}};
        \end{tikzpicture}
    }
    }
}
\tikzstyle{rotBox} = [
  rectangle,
  draw=black,
  minimum width=0.6cm,
  minimum height=0.6cm,
  inner sep=0pt,
  outer sep=0pt,
  align=center,
  text width=0.6cm,
  text height=0.6cm,
  text depth=0cm
]

\newcommand{\dnnSymb}[1]{%
  \makebox[0.8cm][l]{\hspace{-1.05mm}
  \begin{tikzpicture}

    \def\r{0.25mm} 
    \def\xleft{-0.25}
    \def\xmidl{-0.125}
    \def\xmidr{0.125}
    \def\xright{0.25}
    \def\yspace{1.25mm} 
    \def\numNmid{5}
    \def\numNin{3}
    
    \foreach \i in {1,..., \numNin} {
      \node[anchor=center, circle,fill=gray,inner sep=0pt,minimum size=2*\r] (ll\i) at (\xleft,{\yspace*(\i-\numNin+ (\numNmid - \numNin) / 2)}) {};
    }
    \foreach \i in {1,...,\numNmid} {
      \node[anchor=center, circle,fill=gray,inner sep=0pt,minimum size=2*\r] (ml\i) at (\xmidl,{\yspace*(\i-\numNin)}) {};
    }
    \foreach \i in {1,...,\numNmid} {
      \node[anchor=center, circle,fill=gray,inner sep=0pt,minimum size=2*\r] (mr\i) at (\xmidr,{\yspace*(\i-\numNin)}) {};
    }
    \foreach \i in {1,..., \numNin} {
      \node[anchor=center, circle,fill=gray,inner sep=0pt,minimum size=2*\r] (r\i) at (\xright,{\yspace*(\i-\numNin+ (\numNmid - \numNin) / 2)}) {};
    }
    \foreach \li in {1,..., \numNin}{
      \foreach \mi in {1,...,\numNmid}{
        \draw[anchor=center, lightgray,line width=0.25pt] (ll\li) -- (ml\mi);}
    }

    \foreach \li in {1,..., \numNmid}{
      \foreach \mi in {1,...,\numNmid}{
        \draw[anchor=center, lightgray,line width=0.25pt] (ml\li) -- (mr\mi);}
    }
    \foreach \mi in {1,...,\numNmid}{
      \foreach \ri in {1,...,\numNin}{
        \draw[anchor=center, lightgray,line width=0.25pt] (mr\mi) -- (r\ri);}
    }
    \node[anchor=center] at (0, 0) {\contour{white}{#1}};
  \end{tikzpicture}
  }
}
\tikzstyle{dnnBox} = [
  rectangle,
  draw=black,
  minimum width=0.6cm,
  minimum height=0.6cm,
  inner sep=0pt,
  outer sep=0pt,
  align=center,
  text width=0.6cm,
  text height=0.6cm,
  text depth=0cm
]

\tikzstyle{delay} = [
  rectangle,
  draw=black,
  minimum width=0.3cm,
  minimum height=0.3cm,
  inner sep=0pt,
  outer sep=0pt,
  align=center,
  text width=0.3cm,
  text height=0.3cm,
  text depth=0cm,
  fill=white
]
\newcommand{\delaySymb}{
  \makebox[1cm][l]{\hspace{-1mm}\raisebox{0.25mm}{
  \begin{tikzpicture}
  \node {\tiny$z^{\hspace{-0.5mm}\scalebox{0.5}[1.0]{$-$}\hspace{-0.25mm}\scalebox{0.8}{$1$}}$};
  \end{tikzpicture}
  }
  }
}
\tikzstyle{split} = [circle, fill=black, minimum size=\splitSize, inner sep=0pt]
\tikzstyle{add} = [circle, fill=white, minimum size=\addSize, inner sep=0pt, draw=black]
\tikzstyle{arrow} = [line width=\ambixWidth,
    -{Computer Modern Rightarrow[line cap=round,width=6pt,length=2.5pt,line width=1pt]},
    draw=black]

\newcommand{\picLegend}[1]{\raisebox{-0.275ex}{\includegraphics[height=0.75em]{#1}}}  
\newcommand{\picTable}[1]{\raisebox{-0.275ex}{\includegraphics[height=0.9em]{#1}}}  

\newcommand\acknowledgements{This work was supported by the Deutsche Forschungsgemeinschaft (DFG) under Grant 508337379. Computational resources were provided by the Regional Computer Center (RRZ) of the University of Hamburg and the Erlangen National High Performance Computing Center (NHR@FAU) under Project f104ac. NHR is funded by the Federal Government and Bavaria. Both facilities received DFG funding under Grants 440719683 and 498394658.
}

\title{Adaptive Rotary Steering with Joint Autoregression for Robust Extraction of Closely Moving Speakers in Dynamic Scenarios}
\name{Jakob Kienegger, Timo Gerkmann\thanks{\acknowledgements}}
\address{Signal Processing (SP), University of Hamburg, Germany}
\begin{document}
\ninept
\maketitle
\begin{abstract}
Latest advances in deep spatial filtering for Ambisonics demonstrate strong performance in stationary multi-speaker scenarios by rotating the sound field toward a target speaker prior to multi-channel enhancement.
For applicability in dynamic acoustic conditions with moving speakers, we propose to automate this rotary steering using an interleaved tracking algorithm conditioned on the target's initial direction.
However, for nearby or crossing speakers, robust tracking becomes difficult and spatial cues less effective for enhancement.
By incorporating the processed recording as additional guide into both algorithms, our novel joint autoregressive framework leverages temporal-spectral correlations of speech to resolve spatially challenging speaker constellations.
Consequently, our proposed method significantly improves tracking and enhancement of closely spaced speakers, consistently outperforming comparable non-autoregressive methods on a synthetic dataset.
Real-world recordings complement these findings in complex scenarios with multiple speaker crossings and varying speaker-to-array distances.

\end{abstract}
\section{Introduction}
\label{sec:intro}
Speech enhancement is dedicated to improving the clarity and comprehensibility of a recorded speech signal by removing noise and reverberation.
In challenging acoustic environments with multiple talkers, known as the \textit{cocktail party scenario}~\cite{cherry53cocktail_party}, \ac{tse} aims to isolate and enhance the speech of one particular person.
Provided that recordings from a microphone array are available, spatial cues can be used to resolve the inherent ambiguity which speaker to enhance and which to suppress.
If the speakers are distinguishable by their orientation to the array, referred to as \ac{doa}, a \ac{ssf} can be steered toward the target to extract their voice.  
Recent deep, non-linear \acp{ssf} achieve outstanding enhancement performance while retaining computationally efficient \ac{nn} architectures \cite{tesch24ssf_journal, bohlender24sep_journal, pandey12directional_speech_extraction, gu24rezero}.

Spherical microphone arrays are able to capture the sound field in the spherical harmonics representation known as Ambisonics \cite{zotter19ambisonics}. 
The underlying property of rotational invariance enables to re-orient the encoded sound field without physically moving the array.
Instead of an explicit conditioning mechanism to guide a \ac{ssf} toward the target speaker, Wang et al. propose to utilize this feature and center the recording about the desired direction prior to processing \cite{wang25rotary_steering}.
This so-called \textit{rotary steering} is capable of converting widely used but not natively steerable multi-channel enhancement architectures such as \mbox{McNet} \cite{yang23mcnet} or \mbox{SpatialNet} \cite{quan24spatialnet} into \acp{ssf} without any modification.

Recording scenarios such as a seated conference meeting \cite{chen20libricss} may legitimize the assumption that the positions of the speakers remain fixed throughout a recording.
However, in more general settings like the dinner party considered in \cite{barker18chime5}, the movement of the speakers becomes non-negligible. 
To keep using \acp{ssf} for \ac{tse}, either continuous directional information has to be provided, denoted as \textit{strong} guidance, or estimated, e.g., based on the target's initial direction using a tracking algorithm, which we refer to as \textit{weak} guidance \cite{kienegger25wg_ssf, kienegger25sg_ssf}. 
However, when speakers are closely spaced or cross paths, accurate tracking becomes very challenging and the spatial guidance itself loses efficacy in disentangling the speech signals. 

While real-time capability is often a performance-limiting factor for speech enhancement \cite{luo19conv_tasnet, chao24SEmamba}, recent works have capitalized on the enforced sequential processing style.
By incorporating the enhanced recording as additional \ac{nn} input, such \ac{ar} architectures exploit the \textit{temporal-spectral} correlations of speech to achieve improved robustness \cite{andreev23iterative_autoregression, pan24paris_autoregressive_separation, shen25arise}.
Recently, Shen et al. demonstrated the effectiveness of utilizing autoregression for multi-channel speech enhancement in a stationary single-speaker scenario \cite{shen25arise}.

In this work, we generalize rotary steering from stationary to dynamic scenarios by continuously centering the sound field on a moving target speaker.
Using an interleaved tracking algorithm, we automate this adaptive alignment based solely on the target's initial orientation.
Our novel approach leverages rotary steering as a universal conditioning mechanism for both tracking and enhancement and is thereby independent of specific subsystem implementations.
By introducing temporal feedback, we demonstrate on a synthetic three-speaker dataset how the additional guidance from the enhanced signal improves both tracking and enhancement, especially for closely spaced sources. 
As a result, our method outperforms comparable, non-\ac{ar} approaches across different architectures, while maintaining robust enhancement in challenging real-world recordings.

\newcommand\figVspace{-5pt}
\begin{figure*}[t!]
\centering
\hspace*{-5pt}\subcaptionbox{Fixed rotary steering.\vphantom{$\widehat{S}$}\label{fig:fixed_flowchart}}[0.23\linewidth]{%
  \newcommand\nodeDistance{2.5mm}
\newcommand\doubleInputOff{2mm}
\newcommand\specWidth{7mm}
\newcommand\specHeight{5mm}
\newcommand\doaHeight{3.5mm}
\newcommand\spkIdx{1}
\newcommand\condOff{4mm}

\begin{tikzpicture}[node distance=2*\nodeDistance and 1*\nodeDistance]

\node [] (noisy_input) {};
\foreach \x in {0, ..., 3}{
 \pgfmathtruncatemacro{\colorPerc}{round(\x*30 + 15)}
\pgfmathsetlengthmacro{\sOff}{(\x - 1.5) * \shadowOff}
\node [xshift=-\sOff - \specWidth / 2, yshift=-\sOff] at (noisy_input.center) {
  \pgfimage[width=\specWidth, height=\specHeight]{images/flowchart/images/input_spk\spkIdx_\x.pdf}
};
}
\node [anchor=north, xshift=-\specWidth / 2, yshift=-\specHeight / 2] at (noisy_input) {$\mathbf{Y}_t$};

\node [above=of noisy_input.center, xshift=-\specWidth/2] (rot_input) {
\input{images/flowchart/init_doa.tikz}
};

\node [rotBox, right=of noisy_input] (rotate_input_weak) {\rotSymb{$\mathbf{D}^{\top}_0$}};

\node [dnnBox, right=of rotate_input_weak] (ssf)  {\dnnSymb{SSF}};
\node [right=of ssf] (ssf_output) {};
\node [xshift=\specWidth / 2 + 2 * \shadowOff, yshift=-\specHeight] at (ssf_output.west) {$\widehat{S}_t$};
\node [xshift=\specWidth / 2 + 2 * \shadowOff, draw=tab_orange, line width=1pt, inner sep=-0.3pt] at (ssf_output.west) {
  \pgfimage[width=\specWidth, height=\specHeight]{images/flowchart/images/est_spk\spkIdx_0.pdf}
};

\node[anchor=south west, draw=tab_orange, line width=1pt, inner sep=0.4pt] at ($(rotate_input_weak.north) + (3*\shadowOff, 3*\shadowOff + 3mm)$) {
  \pgfimage[width=\specWidth, height=\doaHeight]{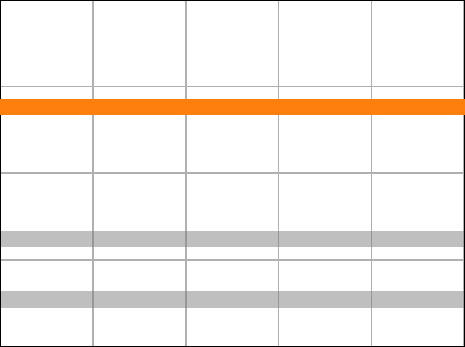}
};

\foreach \x in {0, ..., 3}{
 \pgfmathtruncatemacro{\colorPerc}{round(\x*30 + 15)}
\pgfmathsetlengthmacro{\sOff}{(\x - 1.5) * \shadowOff}
\ifthenelse{\x = 3}{
    \def\shadowColor{black}
}{
    \def\shadowColor{lightgray}
}
\draw [arrow,color=\shadowColor] ($(noisy_input.east) + (-\sOff,-\sOff)$)
        -- ($(rotate_input_weak.west) + (0,-\sOff)$);
\draw [arrow,color=\shadowColor] ($(rotate_input_weak.east) + (0,-\sOff)$)
        -- ($(ssf.west) + (0,-\sOff)$);
}

\draw [arrow] ($(rot_input.east) + (0, \condOff)$) -| (rotate_input_weak);
\draw [arrow] (ssf) -- (ssf_output);

\end{tikzpicture}
\vspace*{\figVspace}
}\hfill
\hspace*{5pt}\subcaptionbox{Adaptive rotary steering utilizing tracking (TST).\vphantom{$\widehat{S}$}\label{fig:adaptive_flowchart}}[0.36\linewidth]{%
  \newcommand\nodeDistance{2.5mm}
\newcommand\doubleInputOff{2mm}
\newcommand\specWidth{7mm}
\newcommand\specHeight{5mm}
\newcommand\doaHeight{3.5mm}
\newcommand\spkIdx{1}
\newcommand\delayOff{2.5mm}
\newcommand\condOff{2mm}
\newcommand\addOff{4mm} %

\begin{tikzpicture}[node distance=2*\nodeDistance and 1*\nodeDistance]

\node [] (noisy_input) {};
\foreach \x in {0, ..., 3}{
 \pgfmathtruncatemacro{\colorPerc}{round(\x*30 + 15)}
\pgfmathsetlengthmacro{\sOff}{(\x - 1.5) * \shadowOff}
\node [xshift=-\sOff - \specWidth / 2, yshift=-\sOff] at (noisy_input.center) {
  \pgfimage[width=\specWidth, height=\specHeight]{images/flowchart/images/input_spk\spkIdx_\x.pdf}
};
}
\node [anchor=north, xshift=-\specWidth / 2, yshift=-\specHeight / 2] at (noisy_input) {$\mathbf{Y}_t$};

\node [right=of noisy_input] (split_tst_input) {};
\node[above=of split_tst_input, yshift=-\delayOff] (tst_delay) {};
\node [rotBox, above right=of tst_delay.center, yshift=-\delayOff] (rotate_input_weak) {\rotSymb{$\mathbf{D}^{\top}_0$}};
\node[dnnBox, right=of rotate_input_weak] (tst) {\dnnSymb{TST}};
\node[add, right=of tst, xshift=\addOff] (tst_add) {\tiny $+$};
\node [rotBox] at (tst_add.center |- noisy_input.center) (rotate_input_strong) {\rotSymb{$\mathbf{\widehat{D}}^{\top}_t$}};

\node[anchor=north west] at ($(tst.east) + (-1*\shadowOff, 1*\shadowOff)$) {\smallMath $\myDelta\widehat{\elevation}_t$
};
\node[anchor=south west] at ($(tst.east) + (-1*\shadowOff, -1*\shadowOff)$) {\smallMath $\myDelta\widehat{\theta}_t$
};
\node[anchor=south west, draw=tab_orange, line width=1pt, inner sep=0.4pt] (init_azimuth) at ($(tst_add.north east) + (0, 1.5mm)$) {
  \pgfimage[width=\specWidth, height=\doaHeight]{images/flowchart/images/ft_jnf-fixed-azimuth.pdf}
};
\node[anchor=south east] at ($(rotate_input_strong.north) + (0, -\shadowOff)$) {\smallMath $\widehat{\theta}_t,\!\widehat{\elevation}_t$
};
\node[anchor=north west, draw=tab_orange, line width=1pt, inner sep=0.4pt] (azimuth) at ($(tst_add.south east) + (0, -1mm)$) {
  \pgfimage[width=\specWidth, height=\doaHeight]{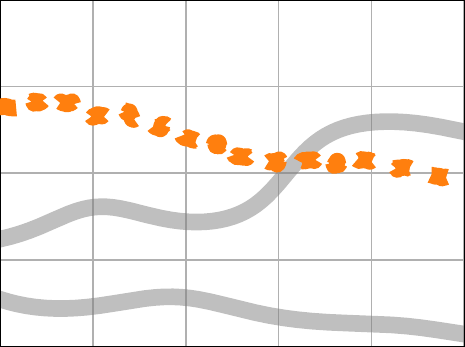}
};

\node [above=of noisy_input.center, xshift=-\specWidth/2] (rot_input) {
\input{images/flowchart/init_doa.tikz}
};
\node [split, above=of rotate_input_weak, yshift=-\condOff] (cond_split) {};

\node [dnnBox, right=of rotate_input_strong] (ssf)  {\dnnSymb{SSF}};
\node [right=of ssf] (ssf_output) {};
\node [xshift=\specWidth / 2 + 2 * \shadowOff, yshift=-\specHeight] at (ssf_output.west) {$\widehat{S}_t$};
\node [xshift=\specWidth / 2 + 2 * \shadowOff, draw=tab_orange, line width=1pt, inner sep=-0.3pt] at (ssf_output.west) {
  \pgfimage[width=\specWidth, height=\specHeight]{images/flowchart/images/est_spk\spkIdx_0.pdf}
};

\foreach \x in {0, ..., 3}{
 \pgfmathtruncatemacro{\colorPerc}{round(\x*30 + 15)}
\pgfmathsetlengthmacro{\sOff}{(\x - 1.5) * \shadowOff}
\ifthenelse{\x = 3}{
    \def\shadowColor{black}
}{
    \def\shadowColor{lightgray}
}
\draw [arrow,color=\shadowColor] ($(noisy_input.east) + (-\sOff,-\sOff)$)
        -- ($(rotate_input_strong.west) + (0,-\sOff)$);
\draw [arrow,color=\shadowColor] ($(rotate_input_weak.east) + (0,-\sOff)$)
        -- ($(tst.west) + (0,-\sOff)$);
\node [split, right=of noisy_input, fill=\shadowColor, yshift=-\sOff, xshift=-\sOff] (split_tst_input_\x) {};
\draw [arrow, draw=\shadowColor] (split_tst_input_\x) |- ($(rotate_input_weak.west) + (0,-\sOff)$);
\draw [arrow,color=\shadowColor] ($(rotate_input_strong.east) + (0,-\sOff)$)
        -- ($(ssf.west) + (0,-\sOff)$);
}

\draw [arrow] (tst) -- (tst_add);
\draw [arrow] (tst_add) -- (rotate_input_strong);
\draw [arrow] (ssf) -- (ssf_output);
\node (cond_input) at (split_tst_input_3.west |- cond_split) {};
\draw (cond_split) -- (cond_input);
\draw[arrow] (cond_split) -- (rotate_input_weak);
\draw[arrow] (cond_split) -| (tst_add);

\end{tikzpicture}
\vspace*{\figVspace}
}\hfill
\subcaptionbox{Integrating the processed signal $\widehat{S}_{\delayIdx{t}}$ as additional guide.\label{fig:ar_flowchart}}[0.40\linewidth]{%
  \newcommand\nodeDistance{2.5mm}
\newcommand\doubleInputOff{1mm}
\newcommand\specWidth{7mm}
\newcommand\specHeight{5mm}
\newcommand\doaHeight{3.5mm}
\newcommand\spkIdx{1}
\newcommand\delayOff{2.5mm}
\newcommand\condOff{2mm}
\newcommand\bendOff{2mm}
\newcommand\addOff{4mm}
\newcommand\doaLabelOff{2mm}
\newcommand\smallDelayIdx[1]{
    \scriptstyle \hspace{-0.25mm}#1\hspace{0.1mm}\raisebox{0.6ex}{\rule{0.3em}{0.06ex}}\hspace{-0.1mm}1
}

\begin{tikzpicture}[node distance=2*\nodeDistance and 1*\nodeDistance]

\node [] (noisy_input) {};
\foreach \x in {0, ..., 3}{
 \pgfmathtruncatemacro{\colorPerc}{round(\x*30 + 15)}
\pgfmathsetlengthmacro{\sOff}{(\x - 1.5) * \shadowOff}
\node [xshift=-\sOff - \specWidth / 2, yshift=-\sOff] at (noisy_input.center) {
  \pgfimage[width=\specWidth, height=\specHeight]{images/flowchart/images/input_spk\spkIdx_\x.pdf}
};
}
\node [anchor=north, xshift=-\specWidth / 2, yshift=-\specHeight / 2] at (noisy_input) {$\mathbf{Y}_t$};

\node [right=of noisy_input] (split_tst_input) {};
\node[above=of split_tst_input, yshift=-\delayOff] (tst_delay) {};
\node [rotBox, above right=of tst_delay.center, yshift=-\delayOff - \doubleInputOff] (rotate_input_weak) {\rotSymb{$\mathbf{D}^{\top}_0$}};
\node[xshift=\bendOff] (tst_input_bend) at (rotate_input_weak.east) {};
\node[dnnBox, anchor=west, xshift=\bendOff, yshift=\doubleInputOff] (tst) at (tst_input_bend) {\dnnSymb{TST}};
\node[add, right=of tst, xshift=\addOff] (tst_add) {\tiny $+$};
\node [rotBox] at (tst_add.center |- noisy_input.center) (rotate_input_strong) {\rotSymb{$\mathbf{\widehat{D}}^{\top}_{\hspace{-0.25mm}\delayIdx{t}}$}};

\node[anchor=south west] at ($(tst) + (\doaLabelOff, -0.5mm)$) {\smallMath $\myDelta\widehat{\theta}_{\smallDelayIdx{t}}$
};
\node[anchor=north west] at ($(tst) + (\doaLabelOff, 0.75mm)$) {\smallMath $\myDelta\widehat{\elevation}_{\smallDelayIdx{t}}$
};
\node[anchor=south east] at ($(rotate_input_strong) + (2*\shadowOff, \doaLabelOff+ 3*\shadowOff)$) {\smallMath $\widehat{\theta}_{\smallDelayIdx{t}}$
};
\node[anchor=south west] at ($(rotate_input_strong) + (-2*\shadowOff, \doaLabelOff+ 3*\shadowOff)$) {\smallMath $\widehat{\elevation}_{\smallDelayIdx{t}}$
};

\node [above=of noisy_input.center, xshift=-\specWidth/2] (rot_input) {
\input{images/flowchart/init_doa.tikz}
};
\node [split, above=of rotate_input_weak, yshift=-\condOff + \doubleInputOff] (cond_split) {};

\node[xshift=\bendOff] (ssf_input_bend) at (rotate_input_strong.east) {};
\node [dnnBox, anchor=west, yshift=\doubleInputOff, xshift=\bendOff] (ssf) at (ssf_input_bend) {\dnnSymb{SSF}};
\node[split, right=of ssf] (split_ssf_output) {};
\node [right=of split_ssf_output] (ssf_output) {};
\node [xshift=\specWidth / 2 + 2 * \shadowOff, yshift=-\specHeight] at (ssf_output.west) {$\widehat{S}_t$};
\node [xshift=\specWidth / 2 + 2 * \shadowOff, draw=tab_orange, line width=1pt, inner sep=-0.3pt] at (ssf_output.west) {
  \pgfimage[width=\specWidth, height=\specHeight]{images/flowchart/images/est_spk\spkIdx_0.pdf}
};

\node[delay, above=of split_ssf_output] (ar_delay) {\delaySymb};
\node[above=of ar_delay, yshift=-1mm] (ar_top) {};
\node[split] (ar_split) at (ar_top -| ssf_input_bend) {};
\node (tst_input_top) at (ar_top -| tst_input_bend) {};

\foreach \x in {0, ..., 3}{
 \pgfmathtruncatemacro{\colorPerc}{round(\x*30 + 15)}
\pgfmathsetlengthmacro{\sOff}{(\x - 1.5) * \shadowOff}
\ifthenelse{\x = 3}{
    \def\shadowColor{black}
}{
    \def\shadowColor{lightgray}
}
\draw [arrow,color=\shadowColor] ($(noisy_input.east) + (-\sOff,-\sOff)$)
        -- ($(rotate_input_strong.west) + (0,-\sOff)$);
\draw [arrow,color=\shadowColor] ($(rotate_input_weak.east) + (0,-\sOff)$)
        -- ($(tst.west) + (0,-\sOff - \doubleInputOff)$);
\node [split, right=of noisy_input, fill=\shadowColor, yshift=-\sOff, xshift=-\sOff] (split_tst_input_\x) {};
\node [delay, above=of split_tst_input_\x, draw=\shadowColor, yshift=-\delayOff] (tst_delay_\x) {\delaySymb};
\draw [draw=\shadowColor] (split_tst_input_\x) -- (tst_delay_\x);
\draw [arrow, draw=\shadowColor] (tst_delay_\x) |- ($(rotate_input_weak.west) + (0,-\sOff)$); 
\draw [arrow,color=\shadowColor] ($(rotate_input_strong.east) + (0,-\sOff)$)
        -- ($(ssf.west) + (0,-\sOff - \doubleInputOff)$);
}

\draw [arrow] (tst) -- (tst_add);
\draw [arrow] (tst_add) -- (rotate_input_strong);
\draw [arrow] (ssf) -- (ssf_output);
\node (cond_input) at (split_tst_input_3.west |- cond_split) {};
\draw (cond_split) -- (cond_input);
\draw[arrow] (cond_split) -- (rotate_input_weak);
\draw[arrow] (cond_split) -| (tst_add);
\draw (split_ssf_output) -- (ar_delay) |- (ar_split);
\draw[arrow] (ar_split) -- (tst_input_top.center) |- ($(tst.west) + (0, \doubleInputOff + 1.5 * \shadowOff)$);
\draw[arrow] (ar_split) |- ($(ssf.west) + (0, \doubleInputOff + 1.5 * \shadowOff)$);

\newcommand\crossLen{0.45pt}
\node (ar_crossing) at (ar_split.center -| tst_add.center) {};
\draw[color=white, rotate=90] ($(ar_crossing.center) + (-\crossLen, \crossLen)$) -- ($(ar_crossing) + (-\crossLen, -\crossLen)$);
\draw[color=white, rotate=90] ($(ar_crossing.center) + (\crossLen, \crossLen)$) -- ($(ar_crossing) + (\crossLen, -\crossLen)$);

\newcommand\switchLen{3mm}
\node (ar_tst) at ($(ar_crossing.center) + (-3.2mm, 0)$) {};
\draw[color=white, line width=2pt] ($(ar_tst.center) + (-0.5*\switchLen, 0)$) -- ($(ar_tst.center) + (0.5*\switchLen,0)$);
\draw ($(ar_tst.center) + (0.5*\switchLen, 0)$) -- ($(ar_tst.center) + (-0.5*\switchLen,0.5*\switchLen)$);
\draw[<-, >={Stealth[scale=0.8]}]
    ($(ar_tst.center) + (-0.15*\switchLen, -0.3*\switchLen)$) arc[start angle=185, end angle=135, radius=\switchLen];
\node[anchor=south west] at (ar_tst) {\scriptsize \contour{white}{AR-TST}};

\node (ar_ssf) at ($(ar_split.center) + (0, -5mm)$) {};
\draw[color=white, line width=2pt] ($(ar_ssf.center) + (0, -0.5*\switchLen)$) -- ($(ar_ssf.center) + (0, 0.5*\switchLen)$);
\draw ($(ar_ssf.center) + (0, 0.5*\switchLen)$) -- ($(ar_ssf.center) + (-0.5*\switchLen,-0.5*\switchLen)$);
\draw[<-, >={Stealth[scale=0.8]}]
    ($(ar_ssf.center) + (0.3*\switchLen, -0.15*\switchLen)$) arc[start angle=275, end angle=225, radius=\switchLen];
\node[anchor=west, yshift=1mm] at (ar_ssf) {\scriptsize \contour{white}{AR-SSF}};

\end{tikzpicture}\hspace*{-5pt}
\vspace*{\figVspace}
\vspace{2pt}
}
\vspace{-5pt}
\caption{
Weakly guided speaker extraction pipelines conditioned on the target's initial \ac{doa} $(\theta_0, \elevation_0)$.
Rotary steering is used to direct the spatial filter (SSF) either constantly toward the initial \ac{doa} (\textit{fixed}) or use tracking (TST) to adjust the steering to the speaker's movement (\textit{adaptive}).
}
\label{fig:flowchart}
\end{figure*}

\section{Problem Setup}
\label{sec:problem}
\subsection{Ambisonics signal model for target speaker extraction}
We assume availability of a recording from a spherical microphone array, capturing the sound field in the spherical harmonics representation known as Ambisonics \cite{zotter19ambisonics}. 
The observation $\mathbf{Y}$ contains the vectorized coefficients according to canonical \ac{acn} \cite{nachbar11ambix}, which commonly correspond to a real-valued spherical harmonics basis $\mathcal{\sphBasis}_\mathrm{N}$ with truncated basis functions at order $n=\mathrm{N}$ \cite{nachbar11ambix}.
In this work, we use the harmonics defined as
\begin{equation}\label{eq:real_sh}
\sphHarmonics_n^m(\theta, \elevation) = N_n^{\left|m\right|} P_n^{\left|m\right|}\left(\sin(\elevation)\right) \begin{cases}
    \sin(\left|m\right|\theta) \,, &m < 0 \\
    \cos(m\,\theta) \,, &m \ge 0
\end{cases}
\end{equation}
for spherical coordinates azimuth $\theta$ and elevation $\elevation$ with associated Legendre functions $P_n^{m}$ \cite{zotter19ambisonics} and $-n \le m \le n$.
Depending on the definition of normalization factor $N_n^{m}$, $\mathcal{\sphBasis}_\mathrm{N}$ is  
either an orthogonal \cite{nachbar11ambix} 
 or orthonormal \cite{su94rotation_real_sh} basis w.r.t. the surface integral inner product on the unit sphere. 
 For analysis, we decompose observation $\mathbf{Y}$ into an additive mixture of Ambisonics coefficients corresponding to anechoic target speech $\mathbf{S}$ and noise $\mathbf{V}$.
Under the assumption of a reverberant, multi-speaker scenario, the latter contains interfering speech signals and reflected components of the target speech.
With $t$ and $k$ denoting frame and frequency bins, this acoustic setup is represented in the \ac{stft} domain as
\begin{equation}\label{eq:signal_model}
    \mathbf{Y}_{tk} = \mathbf{S}_{tk} + \mathbf{V}_{tk} \, .
\end{equation}
In this work we set the goal of reconstructing the monopole coefficient of the anechoic target speech signal $\mathbf{S}_{tk}$, which we indicate as $S_{tk}$.
Thus, the task of \ac{tse} lies in recovering 
$S_{tk}$ from the noisy observations $\mathbf{Y}_{tk}$ conditioned on a speaker-specific characteristic.

\subsection{Rotary steering for stationary speaker extraction}
Spatially selective filters (\acp{ssf}) differentiate the target's speech signal from interfering sources based on spatially discriminative cues. 
In this work, we restrain the cue modality to the target speaker's relative orientation to the array, commonly referred to as \ac{doa}.
In order to inform deep \acp{ssf} of the \ac{doa}, a variety of conditioning mechanisms have been proposed \cite{tesch24ssf_journal, bohlender24sep_journal, pandey12directional_speech_extraction, gu24rezero}. 
Alternatively, when processing in the Ambisonics domain, the properties of the underlying spherical harmonics basis $\mathcal{\sphBasis}_\mathrm{N}$ can be exploited for this purpose. 
Due to closure under rotation in SO(3) for any order N \cite{su94rotation_real_sh, rafaely08beam_steering_wignerD}, the sound field can be aligned toward the target direction by linear transformation as a means of guiding a downstream \ac{ssf}. 
This non-intrusive steering mechanism, introduced by Wang et al. as \textit{rotary steering} \cite{wang25rotary_steering}, is capable of converting arbitrary multi-channel enhancement architectures into \acp{ssf} without any further modification.
The rotational transformation is represented by Wigner-D matrices 
 parametrizable, e.g., using classical Euler angles $\alpha$, $\beta$, $\gamma$.
For a complex-valued spherical harmonics basis $\widetilde{\mathcal{\sphBasis}}_\mathrm{N}$, their entries can be compactly expressed as
 \begin{equation}\label{eq:complex_wignerD}
     \widetilde{D}_{mm'}^n(\alpha, \beta, \gamma) = e^{-\mathrm{j}m\alpha} d_{mm'}^n(\beta) e^{-\mathrm{j}m'\gamma} \, ,
 \end{equation}
with transcendental Wigner-d functions $d_{mm'}^n$ and \texttt{ZYZ} convention for the Euler angles \cite{su94rotation_real_sh, rafaely08beam_steering_wignerD}.
Following \ac{acn} vectorization, these block-diagonal, unitary Wigner-D matrices $\widetilde{\mathbf{D}}$ can be converted via change of basis into their orthogonal real-valued counterparts $\mathbf{D}$ \cite{su94rotation_real_sh} to match the spherical harmonics definition in \cref{eq:real_sh}.
The alignment of the sound field $\mathbf{Y}_{tk}$ used in rotary steering is thus achieved by multiplication with the inverse (i.e., the transpose in this case) of $\mathbf{D}$, representing the orientation of target speaker toward array.
Leveraging the rotational invariance of Ambisonics signals as input conditioning for spatial audio algorithms sets the foundation of our work.

\section{Proposed Method}
\label{sec:method}
\subsection{Adaptive and fixed rotary steering for dynamic scenarios}
Rotary steering was originally introduced for a stationary scenario with static speakers \cite{wang25rotary_steering}. 
In this work, we generalize this concept to dynamic acoustic conditions by continuously aligning the observation $\mathbf{Y}_{tk}$ with the direction of a moving target.
This \textit{adaptive} rotary steering is performed using time-dependent Wigner-D matrices $\mathbf{D}_t$ parameterized by the target’s \ac{doa} $(\theta_t, \elevation_t)$.
Using the \texttt{ZYZ} Euler angle convention, this corresponds to an initial rotation about \texttt{Y} (elevation, $\beta=\elevation_t$) followed by \texttt{Z} (azimuth, $\alpha=\theta_t$), yielding
\begin{equation}\label{eq:adaptiveD}
    \mathbf{D}_t = \mathbf{D}(\theta_t, \elevation_t, 0) \, .
\end{equation}
If continuous ground truth directional guidance is available, adaptive rotary steering can be readily employed to condition a downstream \ac{ssf} for \ac{tse}.
However, in dynamic scenarios with moving speakers, relying on oracle directional cues throughout the recording significantly limits practical applicability. 
As an alternative, weakly guided \ac{tse} uses only the target’s initial \ac{doa} $(\theta_0, \elevation_0)$ \cite{kienegger25wg_ssf, kienegger25sg_ssf}.
Opposed to adaptive rotary steering, we define \textit{fixed} rotary steering as continuously centering the sound field $\mathbf{Y}_{tk}$ at this starting direction using $\mathbf{D}_0$, see \cref{fig:fixed_flowchart}.
This approach provides a one-time spatial cue, allowing the downstream enhancement algorithm to learn the target's temporal-spectral characteristics at the beginning of the recording.

\subsection{Weakly guided adaptive rotary steering framework}
Instead of using temporal-spectral cues throughout the recording, recent works such as \cite{tesch24ssf_journal, bohlender24sep_journal} underline the significant performance potential of spatial guidance under well defined conditions.
Thus, in our own previous work for \ac{tse} but also for the related task of speech separation in \cite{bohlender24sep_journal}, an upstream tracking algorithm is included to estimate the missing temporal evolution of the directional cues and provide continuous guidance to a downstream \ac{ssf}.
However, instead of adapting existing deep \ac{ssl} architectures for \ac{tst}, we propose employing fixed rotary steering by centering the \ac{nn} input at the target's initial direction $(\theta_0, \elevation_0)$. 
As with the \ac{ssf}, this approach can be applied to any \ac{ssl} method without the need for integrating an additional conditioning mechanism.
The alignment with the target's initial direction turns the task of the \ac{tst} algorithm into estimating the time-dependent angular deviation $\big(\Delta\widehat{\theta}_t, \Delta\widehat{\elevation}_t\big)$.
After unification with the starting direction, the downstream \ac{ssf} can now be continuously steered via
\begin{equation}\label{eq:decomposedD}
    \widehat{\mathbf{D}}_t = \mathbf{D}\Big(\theta_0\!+\!\Delta\widehat{\theta}_t, \elevation_0\!+\!\Delta\widehat{\elevation}_t, 0\Big) \, ,
\end{equation}
yielding the weakly guided \ac{tse} pipeline displayed in \cref{fig:adaptive_flowchart}.

\subsection{Additional guidance via temporal-spectral autoregression}
While directionally guided \acp{ssf} achieve outstanding enhancement for sufficiently spaced speakers, as they move closer or cross, the \ac{doa} becomes increasingly difficult to estimate and less discriminative. 
We propose to exploit the \textit{temporal-spectral} pattern of speech to improve the differentiation of speakers under such challenging conditions, which is beneficial for both tracking and enhancement.

\textbf{\textit{AR-SSF}}
Under the assumption of a sequential processing style, the previously enhanced signal $\widehat{S}_{\delayIdx{t}}$ can be utilized as an additional input to a deep speech enhancement architecture. 
Whereas prior work has focused on multi-channel, single-speaker scenarios \cite{shen25arise}, we generalize this approach to \ac{tse}, arguing that the temporal correlation between the spectral structure of $\widehat{S}_{\delayIdx{t}}$ and observation $\mathbf{Y}_t$ is particularly useful to disentangle spatially ambiguous constellations, such as when speakers cross.
\Cref{fig:ar_flowchart} illustrates our proposed \ac{ar} modification of the \ac{ssf}, denoted as \textit{AR-SSF}, in which we concatenate $\widehat{S}_{\delayIdx{t}}$ as an additional channel to the rotated noisy input $\mathbf{Y}_t$.

\textbf{\textit{AR-TST}}
Previous works could demonstrate the potential of using a pre-processed recording for \ac{doa} estimation. 
By extracting the direct-path propagation of the target signal prior to localization, computationally lightweight, statistics-based algorithms achieve dramatic performance improvements \cite{kienegger25sg_ssf, battula25MIMO_localization}.
However, due to continuously centering the sound field in adaptive rotary steering, the target signal's anechoic propagation is trivial and bears no spatial information.
Instead, just as for the AR-SSF, we propose to provide the processed speech signal $\widehat{S}_{\delayIdx{t}}$ as supplementary conditioning to a deep tracking algorithm.
By approximating the rotation matrices as 
\begin{equation}\label{eq:approxD}
    \widehat{\mathbf{D}}_t \approx \widehat{\mathbf{D}}_{\delayIdx{t}} \, ,
\end{equation}
the input modalities $\widehat{S}_{\delayIdx{t}}$ and $\mathbf{Y}_{\delayIdx{t}}$ are aligned in time, 
providing the \ac{tst} algorithm with the matched spectral pattern of the target speech to the noisy recording.
We propose that this helps consistent tracking of the target and minimizes speaker confusions for spatially close interferer.
\Cref{fig:ar_flowchart} displays the resulting \ac{ar} interconnection from processed signal to \ac{tst} algorithm, which we denote as \textit{AR-TST}.

\section{Experimental Setup}
\label{sec:experiments}
\subsection{Dataset}\label{sec:dataset}
\hspace{\the\parindent}\textit{\textbf{Synthetic dataset}}
In order to facilitate \ac{nn} training and evaluation in a controlled acoustic scenario, we create a synthetic dataset consisting of reverberant three-speaker mixtures.
Each simulated recording consists of utterances from the LibriSpeech corpus \cite{panayotov15librispeech} paired according to Libri3Mix \cite{cosentino20librimix}.
We use the image method \cite{allen79image_method} to spatialize the speech signals for shoe-box shaped rooms with reverberation times between 0.2\,s and 0.5\,s.
In particular, our implementation builds upon gpuRIR \cite{diaz18gpu_rir} and computes the Ambisonics coefficients of the image sources by expanding their Green's function into the harmonics in \cref{eq:real_sh} under far-field assumptions \cite{rafaely19spherical_array_processing}.
Together with \ac{acn}, we choose the normalization factors $N^m_n$ to be compliant with the ambiX format \cite{nachbar11ambix} and restrain the simulation to \ac{foa}, thus, $\mathrm{N}=1$.
The movement of the speakers is modeled via randomized, sinusoidal trajectories according to Diaz et al. in \cite{diaz21srp_phat}, which are able to generate diverse motion patterns as demonstrated in 
\cref{fig:trajectories} and our project webpage\textsuperscript{\ref{project_page}}.
At the start, the speakers are separated by at least 15° and their distance to the microphone array remains in the range of 1\,m to 3\,m throughout the recording.
The step-size for the temporal discretization of the trajectories is aligned with the \ac{stft} parametrization, for which we use a square-root Hann window of length $32\,\mathrm{ms}$ and $16\,\mathrm{ms}$ hop-size.

\textit{\textbf{Recorded dataset}}
To test generalizability and robustness in real-world scenarios, we also include recordings with human speakers for evaluation.
We use a \ac{foa} microphone array, which we place centered in a room measuring $9.5\,\mathrm{m}\times 5.1\,\mathrm{m}\times 2.4\,\mathrm{m}$, with a reverberation time of 0.35\,s.
Two male and one female non-native English speakers are instructed to simultaneously read out segments from the Rainbow Passage \cite{fairbanks60rainbow_passage} while randomly moving in the frontal plane of the array, resulting in multiple speaker crossings and significantly varying speaker-to-array distances per recording.
We split the Rainbow Passage into 3 segments of roughly equal length and permute these among all three speakers, yielding 3 recordings of about 30\,s each.
Videos of all recordings can be found on our project page\footnote{\label{project_page}\projectPage}.

\begin{figure}[t!]
\begin{tikzpicture}

\newcommand\doaWidth{1.5cm}
\newcommand\doaHeight{8mm}
\newcommand\doaDist{2mm} %
\newcommand\doaTickVoff{-0.2mm}
\newcommand\doaTickHoff{-0.15mm}
\newcommand\doaLabelDist{3mm}
\newcommand\lrDist{10mm}
\newcommand\tbDist{8mm}
\newcommand\plotStartX{0mm}
\newcommand\plotStartY{0mm}
\newcommand\majorTick{0.75mm}
\newcommand\minorTick{0.5mm}
\newcommand\tickSize{7pt}
\newcommand\tickSkip{7pt} %
\newcommand\timeDist{3.1mm} %
\newcommand\legendTick{4mm}
\newcommand\legendTickTop{-0.25mm}
\newcommand\legendTickBottom{-0.5mm}
\pgfmathsetlengthmacro{\legendTickMiddle}{0.5 * \legendTickTop + 0.5 * \legendTickBottom}
\newcommand\legendOff{1mm}
\newcommand\legendYOff{-2mm} %
\newcommand\centerOff{-1.1cm}
\newcommand\legendWidth{4cm}
\newcommand\legendSkip{4mm}
\newcommand\legendOffProp{2.7cm}
\newcommand\labelWidth{1pt}
\newcommand\descriptX{2.7mm}
\newcommand\descriptY{2.2mm}
\newcommand\legendXBoxOff{3mm}
\newcommand\legendYBoxOff{3.1cm}
\newcommand\legendXBoxWidth{6.4cm}
\newcommand\legendYBoxHeight{4mm}

\foreach \x / \w / \h / \l / \i [
    remember=\dl as \lastLen (initially 0), 
    evaluate=\l as \dl using \l+\lastLen,
]in {
    0/9.1/6.91/7.48/2798,
    1/8.36/6.38/7.85/494,
    2/6.94/6.5/6.45/849,
    3/7.2/9.42/6.6/2409
    }{
    
    \pgfmathsetmacro{\filenum}{int(\i)}
    \def\exampleName{example_\filenum}
    
    \pgfmathsetlengthmacro{\dWidth}{\l / 6 * \doaWidth} %
    \pgfmathsetlengthmacro{\dOff}{\lastLen / 6 * \doaWidth} %
    \pgfmathsetmacro{\dOffVal}{int(abs(\dOff))}
    \pgfmathsetmacro{\dMaxVal}{6 * \doaWidth}
    \node[anchor=mid] at (\plotStartX + \lrDist + \dOff + \x * \doaDist + 0.5 * \dWidth, \plotStartY + 0.5 \doaHeight) {%
        \pgfimage[height=\doaHeight, width=\dWidth]{images/trajectories/\exampleName/azimuth.pdf} 
    };
    \pgfmathsetmacro{\timeNum}{int(\l)}
    \pgfmathsetmacro{\timeOff}{\doaWidth / 6}
    \pgfmathsetmacro{\doaOff}{\doaHeight / 4 - 0.025mm}
    \foreach \t in {0, ..., \timeNum}{
        \pgfmathtruncatemacro\tmp{int(\t/2) * 2}
        \ifnum\tmp=\t
            \pgfmathsetmacro{\tickLen}{\majorTick}
            \pgfmathsetmacro{\time}{int(\t)}
            \node[anchor=north] at (\plotStartX + \lrDist + \dOff + \x * \doaDist + \t * \timeOff, \plotStartY + \doaTickVoff) {\fontsize{\tickSize}{\tickSkip}\selectfont \time};
        \else
            \pgfmathsetmacro{\tickLen}{\minorTick}
        \fi
        \draw[line width=\tickWidth] (\plotStartX + \lrDist + \dOff + \x * \doaDist + \t * \timeOff, \plotStartY + \doaTickVoff) --  (\plotStartX + \lrDist + \dOff + \x * \doaDist + \t * \timeOff, \plotStartY - \tickLen + \doaTickVoff);
    }
    \foreach \d in {0, ..., 4}{
        \pgfmathtruncatemacro\tmp{int(\d/2) * 2}
        \ifnum\tmp=\d
            \pgfmathsetmacro{\tickLen}{\majorTick}
            \ifthenelse{\x = 0}{
                \pgfmathsetmacro{\doa}{int(\d * 90 - 180)}
                \node[anchor=east] at (\plotStartX + \lrDist + \dOff + \x * \doaDist ,\plotStartY + \d * \doaOff + \doaTickHoff) {\fontsize{\tickSize}{\tickSkip}\selectfont \doa};
            }{}
        \else
            \pgfmathsetmacro{\tickLen}{\minorTick}
        \fi
        \draw[line width=\tickWidth] (\plotStartX + \lrDist + \dOff + \x * \doaDist - \tickLen,\plotStartY + \d * \doaOff + \doaTickHoff) --  (\plotStartX + \lrDist + \dOff + \x * \doaDist,\plotStartY + \d * \doaOff + \doaTickHoff);
    }
    
    \pgfmathsetlengthmacro{\W}{\w / 7.5 * \doaWidth} %
    \pgfmathsetlengthmacro{\H}{\h / 7.5 * \doaWidth} %
    \node[anchor=center] at (\plotStartX + \lrDist + \dOff + \x * \doaDist + 0.5 * \W, \plotStartY + 1*\doaHeight + \tbDist + 0.5 * \H) {%
        \pgfimage[height=\H, width=\W]{images/trajectories/\exampleName/room.pdf} 
    };
    \pgfmathsetmacro{\widthNum}{int(\w)}
    \pgfmathsetlengthmacro{\widthOff}{\doaWidth / 7.5}
    \foreach \wn in {0, ..., \widthNum}{
        \pgfmathtruncatemacro\tmp{int(\wn/3) * 3}
        \ifnum\tmp=\wn
            \pgfmathsetmacro{\tickLen}{\majorTick}  
            \node[anchor=north] at ((\plotStartX  + \lrDist + \dOff + \x * \doaDist + \wn * \widthOff, \plotStartY + \doaHeight + \tbDist) {\fontsize{\tickSize}{\tickSkip}\selectfont \wn};
        \else
            \pgfmathsetmacro{\tickLen}{\minorTick}
        \fi
        \draw[line width=\tickWidth] (\plotStartX  + \lrDist + \dOff + \x * \doaDist + \wn * \widthOff, \plotStartY + \doaHeight + \tbDist) --  (\plotStartX  + \lrDist + \dOff + \x * \doaDist + \wn * \widthOff, \plotStartY - \tickLen + \doaHeight + \tbDist);
    }
    \pgfmathsetmacro{\heightNum}{int(\h)}
    \pgfmathsetlengthmacro{\heightOff}{\doaWidth / 7.5}
    \foreach \hn in {0, ..., \heightNum}{
        \pgfmathtruncatemacro\tmp{int(\hn/3) * 3}
        \ifnum\tmp=\hn
            \pgfmathsetmacro{\tickLen}{\majorTick}
            \ifthenelse{\x = 0}{
                \node[anchor=east] at (\plotStartX  + \lrDist + \dOff + \x * \doaDist, \plotStartY + \doaHeight + \tbDist + \hn * \heightOff) {\fontsize{\tickSize}{\tickSkip}\selectfont \hn};
            }{}
        \else
            \pgfmathsetmacro{\tickLen}{\minorTick}
        \fi
        \draw[line width=\tickWidth] (\plotStartX  + \lrDist + \dOff + \x * \doaDist - \tickLen, \plotStartY + \doaHeight + \tbDist + \hn * \heightOff) --  (\plotStartX  + \lrDist + \dOff + \x * \doaDist, \plotStartY + \doaHeight + \tbDist + \hn * \heightOff);
    }
    
}

\node[anchor=north] at (0.5 * \linewidth, \plotStartY- 3*\majorTick + \doaTickVoff) {\footnotesize time [s]};
\node[anchor=north] at (0.5 * \linewidth, \plotStartY- 3*\majorTick + \doaHeight + \tbDist) {\footnotesize room width [m]};
\node[anchor=south, rotate=90] at (\plotStartX  + \lrDist - 4.5mm, \plotStartY + 0.5 * \doaHeight) {\footnotesize DoA\,$\theta_t$\,[°]};
\node[anchor=south, rotate=90] at (\plotStartX  + \lrDist - 2mm, \plotStartY + \doaHeight+ 0.5 * \doaWidth + \tbDist) {\footnotesize length [m]};

\newcommand\labelColor{tab_orange}
\draw[draw, rounded corners=\rectEdge, line width=\tickWidth]
            (\plotStartX + \legendXBoxOff, \plotStartY + \legendYBoxOff) rectangle (\plotStartX + \legendXBoxWidth + \legendXBoxOff, \plotStartY + \legendYBoxOff + \legendYBoxHeight);
            
\newcommand\micLegendDist{4mm}
\node[anchor=west] at (\plotStartX + \legendXBoxOff + \micLegendDist, \plotStartY + \legendYBoxOff+ 0.5 * \legendYBoxHeight) {\footnotesize mic.\,array};
\node[star,star points=5,draw,fill=black,minimum size=5pt,star point ratio=2.25, anchor=center, inner sep=0pt, rounded corners=0.1pt, xshift=-1mm] at (\plotStartX + \legendXBoxOff + \micLegendDist, \plotStartY + \legendYBoxOff+ 0.5 * \legendYBoxHeight) {};

\newcommand\oracleDist{1.875cm}
\draw[color=\labelColor, line width=1.5pt, opacity=0.5,decorate,
  decoration={snake, amplitude=3pt, segment length=10pt}] (\plotStartX + \legendXBoxOff + \oracleDist, \plotStartY + \legendYBoxOff+ 0.5 * \legendYBoxHeight) -- (\plotStartX + \legendXBoxOff + \oracleDist+ \legendSkip, \plotStartY + \legendYBoxOff+ 0.5 * \legendYBoxHeight);
\node[anchor=west, yshift=0.0mm] at (\plotStartX + \legendXBoxOff + \oracleDist+ \legendSkip, \plotStartY + \legendYBoxOff+ 0.5 * \legendYBoxHeight){\footnotesize oracle trajectory};

\newcommand\micStartDist{4.75cm}
\draw (\plotStartX + \legendXBoxOff + \micStartDist-1.75mm, \plotStartY + \legendYBoxOff+ 0.5 * \legendYBoxHeight) 
node[cross=3.5pt, color=\labelColor, line width=1.75pt, anchor=center, rotate=45]{};
\fill[\labelColor, xshift=1.5mm] (\plotStartX + \legendXBoxOff + \micStartDist, \plotStartY + \legendYBoxOff+ 0.5 * \legendYBoxHeight) circle (2pt);
\node[] at (\plotStartX + \legendXBoxOff + \micStartDist, \plotStartY + \legendYBoxOff+ 0.5 * \legendYBoxHeight) {
	\footnotesize $/$
};
\node[xshift=2mm, anchor=west] at (\plotStartX + \legendXBoxOff + \micStartDist, \plotStartY + \legendYBoxOff+ 0.5 * \legendYBoxHeight) {
	\footnotesize start$/$stop
};

\end{tikzpicture}
\vspace*{-17.5pt}
\caption{
Synthetic three-speaker ({\protect\tikz[baseline=-1.2ex]  \protect\draw[tab_blue, line width=1pt] (0,0mm) -- (1mm,0mm);}/{\protect\tikz[baseline=-0.7ex] \protect\draw[tab_orange, line width=1pt] (0,0mm) -- (1mm,0mm);}/{\protect\tikz[baseline=-0.2ex] \protect\draw[tab_green, line width=1pt] (0,0mm) -- (1mm,0mm);}) dataset modeling continuous motion \cite{diaz21srp_phat}. Tracking performance is shown for azimuth \ac{doa} $\theta_t$ using non-AR 
({\protect\tikz[baseline=-0.6ex]  \protect\draw[color=tab_orange, line width=1pt, decorate, dash pattern=on 1pt off 1pt,
  decoration={snake, amplitude=3pt, segment length=10pt}] (0, 0) -- (4mm, 0);}) and our AR modification ({\protect\tikz[baseline=-0.6ex]  \protect\draw[color=tab_orange, line width=1pt, decorate, dash pattern=on 3pt off 1pt,
  decoration={snake, amplitude=3pt, segment length=10pt}] (0, 0) -- (4mm, 0);}) of SELDnet~\cite{yasuda24causal_seldnet}.
}
\label{fig:trajectories}
\vspace*{-5pt}
\end{figure}

\subsection{Algorithm implementation and optimization details}\label{sec:nn_training}

\hspace{\the\parindent}\textit{\textbf{Spatially selective filter}}
We demonstrate the generalizability of our methods using the two established multi-channel enhancement models \mbox{McNet} \cite{yang23mcnet} and \mbox{SpatialNet} \cite{quan24spatialnet}.
Although lacking directional conditioning, rotary steering enables their use as \acp{ssf} by applying a network-agnostic pre-processing step to the noisy Ambisonics input.
 We adopt causal variants of both, employing the Mamba state-space models for narrow-band processing in \mbox{SpatialNet} \cite{quan24online_spatialnet}.
\Cref{tab:results} compares model size and computational cost measured in \acp{mac} between the two \acp{nn}.
In both cases, the additional input channel for the AR-SSFs only lead to a minimal increase of less than 1\,\% in model parameters and 1.5\,\% in \acp{mac}.

\textit{\textbf{Target speaker tracking}}
We choose the convolutional-recurrent architecture SELDnet \cite{adavanne19seldnet} as baseline model for \ac{tst}. 
Originally proposed for the composite \ac{seld} task, %
we adapt it for our setup by removing the sound event detection output and employing the causal implementation from \cite{yasuda24causal_seldnet}. 
The resulting lightweight model has less than 300\,k parameters and 70\,M/s \acp{mac} for both non-\ac{ar} and \ac{ar} versions.  
 
\textit{\textbf{Training strategy}}
To guarantee a robust interplay between tracking and enhancement algorithms, a joint training approach is necessary \cite{kienegger25wg_ssf, kienegger25sg_ssf}. 
Instead of fine-tuning, we use a dual-optimization strategy with a single forward pass through the full pipeline, followed by backpropagation via individual optimizers for \ac{ssf} and \ac{tst}.
To avoid the inherent non-parallelizability of the \ac{ar} methods, we use \acl{rds} as pseudo-\ac{ar} training framework \cite{shen25arise}.
We adhere to the original loss functions and learning rate schedulers for optimizing \mbox{McNet}~\cite{yang23mcnet}, \mbox{SpatialNet}~\cite{quan24online_spatialnet} and SELDnet~\cite{adavanne19seldnet}.
Training runs for up to 100 epochs or until convergence, which we define as no improvement for 10 consecutive epochs with either optimizer.

\section{Results}
\label{sec:results}
\newcommand\tableW{3pt}
\newcommand\tableWL{8pt}
\newcommand\tableHspacer{-0pt} %
\newcommand\tableHrow{-0pt} %

\begin{table}[t]
\vspace*{5pt}
\resizebox{\linewidth}{!}{
\renewcommand{\arraystretch}{0.5}
\footnotesize
\begin{tabular}{
@{\hspace{1pt}}c@{\hspace{\tableWL}}l@{\hspace{\tableW}}c@{\hspace{\tableW}}c@{\hspace{\tableW}}c@{\hspace{\tableW}}c@{\hspace{\tableWL}}c@{\hspace{\tableW}}c@{\hspace{1pt}}
}
 \toprule[1.5pt]
 & \multicolumn{5}{c}{\textbf{Pipeline}\,[{\scriptsize McNet$/$SpatialNet}]} & \multicolumn{2}{c}{\textbf{Performance}}\\[1pt] \cmidrule(l{0pt}r{8pt}){2-6} \cmidrule(l{0pt}r{2pt}){7-8}
\textbf{ID} & Tracking & AR-SSF & AR-TST & Params\,[M] & MACs\,[G/s] & PESQ\,$\uparrow$ & ESTOI\,[\%]\,$\uparrow$ \\
\midrule
\noalign{\vskip \tableHspacer}
\picTable{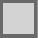} & \multicolumn{1}{c}{$-$} & $-$ & $-$ & $-$ & $-$ & 1.08 & 31.6 \\ \midrule
\noalign{\vskip \tableHspacer}
 \picTable{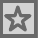} &  Oracle & {\color{red}\ding{55}} & $-$ & 1.84$/$1.74  & 29.9$/$18.7 & 2.08$/$2.27 & 77.7$/$83.0  \\[\tableHrow]
\picTable{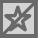} & Oracle & {\color{green}\ding{51}} & $-$ & 1.86$/$1.74 &  30.3$/$18.8 & \textbf{2.21}$/$\textbf{2.38} & \textbf{80.4}$/$\textbf{84.7} \\ \midrule
\noalign{\vskip \tableHspacer}
\picTable{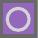} & \multicolumn{1}{c}{$-$} & {\color{red}\ding{55}} & $-$ & 1.84$/$1.74 & 29.9$/$18.7 & 1.97$/$2.11 & 75.1$/$80.3\\[\tableHrow]
\picTable{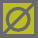} & \multicolumn{1}{c}{$-$}& {\color{green}\ding{51}} & $-$ & 1.86$/$1.74  & 30.3$/$18.8 & \textbf{2.08}$/$\textbf{2.21} & \textbf{78.5}$/$\textbf{82.2}\\ \midrule
\noalign{\vskip \tableHspacer}
\picTable{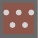} & SELDnet & {\color{red}\ding{55}} & {\color{red}\ding{55}} & 2.13$/$2.03 & 30.0$/$18.8 & 1.98$/$2.11 & 75.5$/$80.6\\[\tableHrow]
\picTable{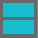} & SELDnet & {\color{red}\ding{55}} & {\color{green}\ding{51}} & 2.13$/$2.03 & 30.0$/$18.8 & 2.03$/$2.18 & 76.6$/$81.6\\ [\tableHrow]
\picTable{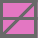} & SELDnet & {\color{green}\ding{51}} & {\color{green}\ding{51}} & 2.15$/$2.03 & 30.3$/$18.8 & \textbf{2.17}$/$\textbf{2.27} & \textbf{79.5}$/$\textbf{82.7}\\ 
 \bottomrule[1.5pt]
\end{tabular}
}
\vspace*{-5pt}
 \caption{
 Model complexity and performance on our synthetic data-set for various speaker extraction approaches and implementations. 
}
\label{tab:results}
\vspace*{-5pt}
\end{table}

\begin{figure}[b!]
\vspace*{-10pt} 
\begin{tikzpicture}
    
\newcommand\ssfNum{3}
\newcommand\ssfName{mcnet}
\newcommand\doaWidth{1.5cm}
\newcommand\doaHeight{8mm}
\newcommand\figSize{2.0cm}
\newcommand\labelFont{\footnotesize}
\pgfmathsetlengthmacro{\figWidth}{\figSize * 1.5}
\pgfmathsetlengthmacro{\figHeight}{\figSize * 0.9}
\newcommand\labelOff{8mm}
\newcommand\doaDist{2cm} %
\newcommand\doaTickVoff{-0.2mm}
\newcommand\doaTickHoff{-0.15mm}
\newcommand\doaLabelDist{3mm}
\newcommand\lrDist{10mm}
\newcommand\tbDist{8mm}
\newcommand\plotStartX{0mm}
\newcommand\plotStartY{0mm}
\newcommand\majorTick{0.75mm}
\newcommand\minorTick{0.5mm}
\newcommand\tickSize{7pt}
\newcommand\tickSkip{9pt}
\newcommand\timeDist{3.1mm} %
\newcommand\legendTick{4mm}
\newcommand\legendTickTop{-0.25mm}
\newcommand\legendTickBottom{-0.5mm}
\pgfmathsetlengthmacro{\legendTickMiddle}{0.5 * \legendTickTop + 0.5 * \legendTickBottom}
\newcommand\legendOff{1mm}
\newcommand\legendYOff{-2mm} %
\newcommand\centerOff{-1.1cm}
\newcommand\legendWidth{4cm}
\newcommand\legendSkip{4mm}
\newcommand\legendOffProp{2.7cm}
\newcommand\labelWidth{1pt}
\newcommand\descriptX{2.7mm}
\newcommand\descriptY{2.2mm}
\newcommand\legendXBoxOff{1.25cm}
\newcommand\legendYBoxOff{4mm}
\newcommand\legendXBoxWidth{2.8cm}
\newcommand\legendYBoxHeight{2.05cm}

\foreach \x / \low / \high / \metric / \numTicks / \metricLabel / \relWidth 
[
    remember=\dr as \lastRel (initially 0), 
    evaluate=\relWidth as \dr using \relWidth+\lastRel,
]
in {
    0/0/10/mae/5/{\small MAE\,[{\scriptsize $^\circ$}]\,$\leftarrow$}/1.25,
    1/1.4/2.4/pesq/5/{\small PESQ\,$\rightarrow$}/1.9
    }{
    
    \pgfmathsetlengthmacro{\figWidth}{\figSize * \relWidth}
    \node[anchor=center] at (\plotStartX + \x * \doaDist + \lastRel * \figSize, \plotStartY + 0.5 * \figHeight) {%
        \pgfimage[height=\figHeight, width=\figWidth]{images/min_angle/mcnet-min_angle_\metric.pdf} 
    };

    \ifnum\x=0
    \node[anchor=center] at (\plotStartX + 2.3cm, \plotStartY - 7mm) {\small Average angular distance to closest interfering speaker [{\scriptsize $^\circ$}]};
    \fi
    \ifthenelse{\x = 0}{
        \foreach \relLabelX / \labelX in {
    0.06 / {\labelFont [0,\,15)}, 0.41 / {\labelFont[45,\,90)}, 0.87 / {\labelFont [135,\,180]}
    }{
    \draw[line width=\tickWidth] (\plotStartX + \lastRel * \figSize + \x * \doaDist - 0.5 * \figWidth + \relLabelX * \figWidth,\plotStartY) --  (\plotStartX + \lastRel * \figSize + \x * \doaDist - 0.5 * \figWidth + \relLabelX * \figWidth,\plotStartY - \majorTick);
    \node[anchor=north] at (\plotStartX + \lastRel * \figSize + \x * \doaDist - 0.5 * \figWidth + \relLabelX * \figWidth,\plotStartY) {\labelX};
    }
    }{
        \foreach \relLabelX / \labelX in {
    0.05 / {\labelFont [0,\,15)}, 0.4 / {\labelFont[45,\,90)}, 0.875 / {\labelFont [135,\,180]}
    }{
    \draw[line width=\tickWidth] (\plotStartX + \lastRel * \figSize + \x * \doaDist - 0.5 * \figWidth + \relLabelX * \figWidth,\plotStartY) --  (\plotStartX + \lastRel * \figSize + \x * \doaDist - 0.5 * \figWidth + \relLabelX * \figWidth,\plotStartY - \majorTick);
    \node[anchor=north] at (\plotStartX + \lastRel * \figSize + \x * \doaDist - 0.5 * \figWidth + \relLabelX * \figWidth,\plotStartY) {\labelX};
    }
    }

    \pgfmathsetlengthmacro{\deltaY}{\figHeight / \numTicks}
    \foreach \y in {0, ..., \numTicks}{
        \pgfmathtruncatemacro\tmp{int(\y/2) * 2}
        \ifthenelse{\y = \numTicks}{
            \newcommand\metricTick{\high}
        }{
            \pgfmathsetmacro{\metricTick}{
        round((\low + (\high - \low) / \numTicks * \y) * 100) / 100
        }
        }
        \ifthenelse{\equal{\metric}{mae}}{
        \ifnum\y=0
\node[anchor=center, rotate=90] at (\plotStartX + \lastRel * \figSize + \x * \doaDist - 0.5 * \figWidth - \labelOff + 2mm,\plotStartY + 0.5 * \figHeight) {\metricLabel};
        \fi        
        \pgfmathsetmacro{\metricTick}{int((\high - \low) / \numTicks * \y)}
         \pgfmathsetmacro{\tickLen}{\majorTick}
            \node[anchor=east] at (\plotStartX + \lastRel * \figSize + \x * \doaDist - \tickLen - 0.5 * \figWidth,\plotStartY + \y * \deltaY) {\labelFont \metricTick};
        }{
            \ifnum\y=0
            \node[anchor=center, rotate=90] at (\plotStartX + \lastRel * \figSize + \x * \doaDist - 0.5 * \figWidth - \labelOff,\plotStartY + 0.5 * \figHeight) {\small \metricLabel};
            \fi
            \pgfmathsetmacro{\tickLen}{\majorTick}
            \node[anchor=east] at (\plotStartX + \lastRel * \figSize + \x * \doaDist - \tickLen - 0.5 * \figWidth,\plotStartY + \y * \deltaY) {\labelFont \num[round-mode=places,round-precision=1, mode=text]{\metricTick}};
        }
        \draw[line width=\tickWidth] (\plotStartX + \lastRel * \figSize + \x * \doaDist - \tickLen - 0.5 * \figWidth,\plotStartY + \y * \deltaY) --  (\plotStartX + \lastRel * \figSize + \x * \doaDist - 0.5 * \figWidth,\plotStartY + \y * \deltaY);
    }
}

\end{tikzpicture}
\vspace*{-18.5pt}
\caption{
Tracking (\acs{mae}) and enhancement (\acs{pesq}) performance dependency of McNet on the distance to the closest interfering speaker.
We report the sample mean with 95\% confidence interval error bars.
}
\vspace*{-3.5pt}
\label{fig:min_angle}
\end{figure}

We use our synthetic dataset for a detailed \textit{analysis} and lab recordings to assess real-world \textit{generalizability}. In both cases, we focus on perceptual quality and intelligibility as key performance measures.

\textbf{\textit{Synthetic dataset}}
With paired ground truth anechoic speech signals, we use the standard intrusive metrics \mbox{PESQ}~\cite{rix01pesq} and \mbox{ESTOI}~\cite{jensen16estoi} to measure speech quality and intelligibility, respectively.
\Cref{tab:results} and \cref{fig:strong_weak} summarize the speaker extraction results for all evaluated  methods, with the strongly guided serving as an upper bound for their weakly guided counterparts.
Both \mbox{McNet} and \mbox{SpatialNet} demonstrate the effectiveness of continuous ground truth directional conditioning (\picLegend{images/legend/strong.pdf},\,\picLegend{images/legend/strong-ar-spectral.pdf}), achieving significant improvements over solely using the initial \ac{doa} (\picLegend{images/legend/weak.pdf},\,\picLegend{images/legend/weak-ar-spectral.pdf}).
However, guidance with \mbox{SELDnet} as tracking algorithm~(\picLegend{images/legend/weak-seldnet.pdf}) yields only marginal benefits over fixed rotary steering~(\picLegend{images/legend/weak.pdf}). 
As shown in \cref{fig:min_angle} (\picLegend{images/legend/weak-seldnet.pdf}) with utterance-wise \ac{mae}, \mbox{SELDnet} performs especially poorly for closely spaced speakers, likely due to limited model capacity.
Although resulting in only a negligible increase in complexity, the \ac{ar} integration of the enhanced speech using our proposed AR-TST framework substantially improves tracking of \mbox{SELDnet} for nearby speakers, see \cref{fig:min_angle} (\picLegend{images/legend/weak-seldnet-ar-spatial.pdf}) complemented by the sample trajectories in \cref{fig:trajectories} and our project page\textsuperscript{\ref{project_page}}.
Consequently, the downstream \ac{ssf} is able to leverage directional guidance to a greater extent, facilitating superior extraction quality.
By incorporating the processed speech signal as input to the \ac{ssf}, AR-SSF (\picLegend{images/legend/weak-ar-spectral.pdf}) yields even greater improvements over tracking-based methods (\picLegend{images/legend/weak-seldnet.pdf},\,\picLegend{images/legend/weak-seldnet-ar-spatial.pdf}) in these scenarios.
This is most evident for McNet, suggesting that input concatenation at each sublayer \cite{yang23mcnet} strongly facilitates temporal-spectral cue integration.
The results for both AR-TST and AR-SSF underline the critical role of spectral cues in spatially challenging speaker scenarios.
Combining these approaches into a novel, \textit{joint} \ac{ar} framework (\picLegend{images/legend/weak-seldnet-ar-spatial-spectral.pdf}) yields consistent improvements across all other weakly guided methods (\picLegend{images/legend/weak.pdf},\,\picLegend{images/legend/weak-ar-spectral.pdf},\,\picLegend{images/legend/weak-seldnet.pdf},\,\picLegend{images/legend/weak-seldnet-ar-spatial.pdf}).
As a result, our proposed joint \ac{ar} approach (\picLegend{images/legend/weak-seldnet-ar-spatial-spectral.pdf}) achieves parity for \mbox{SpatialNet} and substantial improvements for \mbox{McNet} over their non-\ac{ar}, but strongly guided counterparts (\picLegend{images/legend/strong.pdf}).

\begin{figure}[t!]
\begin{tikzpicture}
    
\newcommand\ssfNum{2} %
\newcommand\doaWidth{1.5cm}
\newcommand\doaHeight{8mm}
\newcommand\figSize{1.8cm}
\pgfmathsetlengthmacro{\figWidth}{\figSize * 1.7}
\pgfmathsetlengthmacro{\figHeight}{\figSize}
\newcommand\labelOff{8mm}
\newcommand\labelFont{\footnotesize}
\newcommand\doaDist{1.3cm} %
\newcommand\doaTickVoff{-0.2mm}
\newcommand\doaTickHoff{-0.15mm}
\newcommand\doaLabelDist{3mm}
\newcommand\lrDist{10mm}
\newcommand\tbDist{8mm}
\newcommand\plotStartX{0mm}
\newcommand\plotStartY{0mm}
\newcommand\majorTick{0.75mm}
\newcommand\minorTick{0.5mm}
\newcommand\tickSize{7pt}
\newcommand\tickSkip{9pt}
\newcommand\timeDist{3.1mm} %
\newcommand\legendTick{4mm}
\newcommand\legendTickTop{-0.25mm}
\newcommand\legendTickBottom{-0.5mm}
\pgfmathsetlengthmacro{\legendTickMiddle}{0.5 * \legendTickTop + 0.5 * \legendTickBottom}
\newcommand\legendOff{1mm}
\newcommand\legendYOff{-2mm} %
\newcommand\centerOff{-1.1cm}
\newcommand\legendWidth{4cm}
\newcommand\legendSkip{4mm}
\newcommand\legendOffProp{2.7cm}
\newcommand\labelWidth{1pt}
\newcommand\descriptX{2.7mm}
\newcommand\descriptY{2.2mm}
\newcommand\legendXBoxOff{1.25cm}
\newcommand\legendYBoxOff{4mm}
\newcommand\legendXBoxWidth{2.8cm}
\newcommand\legendYBoxHeight{2.05cm}

\foreach \x / \low / \high / \l / \metric / \numTicks / \metricLabel [
    remember=\dl as \lastLen (initially 0), 
    evaluate=\l as \dl using \l+\lastLen,
]in {
    0/0.8/1.4/7.48/pesq/6/{$\Delta$PESQ\,$\rightarrow$},
    1/0.4/0.54/7.85/estoi/7/{$\Delta$ESTOI\,[{\scriptsize \%}]\,$\rightarrow$}
    }{
    
    \node[anchor=center] at (\plotStartX + \x * \figWidth + \x * \doaDist, \plotStartY + 0.5 * \figHeight) {%
        \pgfimage[height=\figHeight, width=\figWidth]{images/strong_weak/\metric.pdf} 
    };

    \pgfmathsetlengthmacro{\ssfDelta}{\figWidth / \ssfNum}
    \foreach \ssfIdx / \ssfName in {
        0/{McNet}, 1/{SpatialNet}
        }{
        \draw[line width=\tickWidth] (\plotStartX + \x * \figWidth + \x * \doaDist + 0.5 * \ssfDelta + \ssfIdx * \ssfDelta - 0.5 * \figWidth,\plotStartY) --  (\plotStartX + \x * \figWidth + \x * \doaDist + 0.5 * \ssfDelta + \ssfIdx * \ssfDelta - 0.5 * \figWidth,\plotStartY - \majorTick);
        \node[anchor=north, rotate=0] at (\plotStartX + \x * \figWidth + \x * \doaDist + 0.5 * \ssfDelta + \ssfIdx * \ssfDelta - 0.5 * \figWidth,\plotStartY) {\small \ssfName};
    }

    \node[anchor=center, rotate=90] at (\plotStartX + \x * \figWidth + \x * \doaDist - 0.5 * \figWidth - \labelOff,\plotStartY + 0.5 * \figHeight) {\small \metricLabel};
    
    \pgfmathsetlengthmacro{\deltaY}{\figHeight / \numTicks}
    \foreach \y in {0, ..., \numTicks}{
        \ifthenelse{\y = \numTicks}{
            \newcommand\metricTick{\high}
        }{
            \pgfmathsetmacro{\metricTick}{
        round((\low + (\high - \low) / \numTicks * \y) * 100) / 100
        }
        }
        \pgfmathtruncatemacro\tmp{int(\y/2) * 2}
        \ifnum\tmp=\y
        \pgfmathsetmacro{\tickLen}{\majorTick}
        \ifthenelse{\equal{\metric}{estoi}}{
        \pgfmathsetmacro{\metricTick}{int((\high - \low) / \numTicks * \y * 100 + 100 * \low + 1)}
            \node[anchor=east] at (\plotStartX + \x * \figWidth + \x * \doaDist - \tickLen - 0.5 * \figWidth,\plotStartY + \y * \deltaY) {\labelFont \metricTick};
        }{\node[anchor=east] at (\plotStartX + \x * \figWidth + \x * \doaDist - \tickLen - 0.5 * \figWidth,\plotStartY + \y * \deltaY) {\labelFont \num[round-mode=places,round-precision=1, mode=text]{\metricTick}};}
        \else
            \pgfmathsetmacro{\tickLen}{\minorTick}
        \fi
        \draw[line width=\tickWidth] (\plotStartX + \x * \figWidth + \x * \doaDist - \tickLen - 0.5 * \figWidth,\plotStartY + \y * \deltaY) --  (\plotStartX + \x * \figWidth + \x * \doaDist - 0.5 * \figWidth,\plotStartY + \y * \deltaY);
    }
}

\end{tikzpicture}
\vspace*{-20pt}
\caption{
Strongly guided  (\picLegend{images/legend/strong.pdf},\,\picLegend{images/legend/strong-ar-spectral.pdf}), fixed  (\picLegend{images/legend/weak.pdf},\,\picLegend{images/legend/weak-ar-spectral.pdf}) and adaptive (\picLegend{images/legend/weak-seldnet.pdf},\,\picLegend{images/legend/weak-seldnet-ar-spatial.pdf},\,\picLegend{images/legend/weak-seldnet-ar-spatial-spectral.pdf}) weakly guided extraction using different enhancement algorithms.
We report the sample mean with 95\% confidence interval error bars.
}
\label{fig:strong_weak}
\vspace*{-10pt}
\end{figure}

\textbf{\textit{Recorded dataset}}
Due to the lack of paired ground truth speech for our lab recordings, we employ a non-intrusive metric to assess perceptual enhancement quality. 
In particular, we choose \acs{nisqa}~\cite{mittag21nisqa}, a data-driven estimator of the subjective \acl{mos}, which is trained on labeled datasets with human listening test results.
As a measure of intelligibility, we leverage the transcription of a downstream \ac{asr} system to compute the \ac{wer} relative to the Rainbow Passage reference segments.
We employ the lightweight \ac{asr} model \texttt{QuartzNet15x5Base-En} from the NeMo toolkit~\cite{kuchaiev9nemo}, which is trained solely on clean and telephony conversational English speech and therefore particularly sensitive to signal distortions. %
\Cref{fig:recordings} presents the enhancement results in terms of perceptual quality (NISQA) and intelligibility (WER) using weakly guided speaker extraction methods employing adaptive rotary steering. 
Overall, the lab recordings complement the trends observed on the synthetic dataset, showing clear improvements from the non-\ac{ar} approach (\picLegend{images/legend/weak-seldnet}), through the integration of AR-TST (\picLegend{images/legend/weak-seldnet-ar-spatial}), to the combined joint \ac{ar} pipeline (\picLegend{images/legend/weak-seldnet-ar-spatial-spectral}). 
Listening tests, which are accessible on our project page\textsuperscript{\ref{project_page}}, reveal that the performance differences are especially evident in the latter portion of the 30\,s samples. 
While our proposed joint \ac{ar} pipeline (\picLegend{images/legend/weak-seldnet-ar-spatial-spectral}) robustly recovers from speaker crossings and large signal-to-noise variations, the non-\ac{ar} approach (\picLegend{images/legend/weak-seldnet}) frequently loses the target speaker, resulting in spectral leakage and speaker confusions. 
Thus, by deliberately including a large number of crossings and consistently close distances between speakers, the benefits from joint \textit{temporal-spectral} guidance become particularly apparent.

\vspace*{-5pt}
\begin{figure}[htb]
\begin{tikzpicture}

\newcommand\doaWidth{1.5cm}
\newcommand\doaHeight{8mm}
\newcommand\figSize{1.8cm}
\pgfmathsetlengthmacro{\figWidth}{\figSize * 1.7}
\pgfmathsetlengthmacro{\figHeight}{\figSize}
\newcommand\labelOff{8mm}
\newcommand\labelFont{\footnotesize}
\newcommand\doaDist{1.3cm} %
\newcommand\doaTickVoff{-0.2mm}
\newcommand\doaTickHoff{-0.15mm}
\newcommand\doaLabelDist{3mm}
\newcommand\lrDist{10mm}
\newcommand\tbDist{8mm}
\newcommand\plotStartX{0mm}
\newcommand\plotStartY{0mm}
\newcommand\majorTick{0.75mm}
\newcommand\minorTick{0.5mm}
\newcommand\tickSize{7pt}
\newcommand\tickSkip{9pt}
\newcommand\timeDist{3.1mm} %
\newcommand\legendTick{4mm}
\newcommand\legendTickTop{-0.25mm}
\newcommand\legendTickBottom{-0.5mm}
\pgfmathsetlengthmacro{\legendTickMiddle}{0.5 * \legendTickTop + 0.5 * \legendTickBottom}
\newcommand\legendOff{1mm}
\newcommand\legendYOff{-2mm} %
\newcommand\centerOff{-1.1cm}
\newcommand\legendWidth{4cm}
\newcommand\legendSkip{4mm}
\newcommand\legendOffProp{2.7cm}
\newcommand\labelWidth{1pt}
\newcommand\descriptX{2.7mm}
\newcommand\descriptY{2.2mm}
\newcommand\legendXBoxOff{1.25cm}
\newcommand\legendYBoxOff{4mm}
\newcommand\legendXBoxWidth{2.8cm}
\newcommand\legendYBoxHeight{2.05cm}

\foreach \x / \low / \high / \l / \metric / \numTicks / \metricLabel [
    remember=\dl as \lastLen (initially 0), 
    evaluate=\l as \dl using \l+\lastLen,
]in {
    1/1.0/4.5/7.85/nisqa/7/{NISQA\,$\rightarrow$},
    2/0/1.125/6.45/wer/9/{WER\,[{\scriptsize\%}]\,$\leftarrow$}
    }{
    \edef\fileName{\metric}
    
    \node[anchor=center] at (\plotStartX + \x * \figWidth + \x * \doaDist, \plotStartY + 0.5 * \figHeight) {%
        \pgfimage[height=\figHeight, width=\figWidth]{images/recordings/\fileName.pdf} 
    };

    \node[anchor=center, rotate=90] at (\plotStartX + \x * \figWidth + \x * \doaDist - 0.5 * \figWidth - \labelOff,\plotStartY + 0.5 * \figHeight) {\small \metricLabel};
    
    \pgfmathsetlengthmacro{\deltaY}{\figHeight / \numTicks}
    \foreach \y in {0, ..., \numTicks}{
        \pgfmathtruncatemacro\tmp{int(\y/2) * 2}
        \ifthenelse{\y = \numTicks}{
            \newcommand\metricTick{\high}
        }{
            \pgfmathsetmacro{\metricTick}{
        (\low + (\high - \low) / \numTicks * \y)
        }
        }
        \ifthenelse{\equal{\metric}{wer}}{
        \pgfmathsetmacro{\metricTick}{int(12.5 * \y)}
        \ifnum\tmp=\y
         \pgfmathsetmacro{\tickLen}{\majorTick}
            \node[anchor=east] at (\plotStartX + \x * \figWidth + \x * \doaDist - \tickLen - 0.5 * \figWidth,\plotStartY + \y * \deltaY) {\labelFont \text{\metricTick}};
        \else
            \pgfmathsetmacro{\tickLen}{\minorTick}
        \fi
        }{
            \ifnum\tmp=\y
            \pgfmathsetmacro{\tickLen}{\majorTick}
            \node[anchor=east] at (\plotStartX + \x * \figWidth + \x * \doaDist - \tickLen - 0.5 * \figWidth,\plotStartY + \y * \deltaY) {\labelFont \text{\metricTick}};
        \else
            \pgfmathsetmacro{\tickLen}{\minorTick}
        \fi
        }
        \draw[line width=\tickWidth] (\plotStartX + \x * \figWidth + \x * \doaDist - \tickLen - 0.5 * \figWidth,\plotStartY + \y * \deltaY) --  (\plotStartX + \x * \figWidth + \x * \doaDist - 0.5 * \figWidth,\plotStartY + \y * \deltaY);
    }

    \foreach \xTickOff / \xLabel in {
    0.1/{$-$}, 0.4/{McNet}, 0.8/{SpatialNet}
    }{
        \draw[line width=\tickWidth] (\plotStartX + \x * \figWidth + \xTickOff * \figWidth + \x * \doaDist - 0.5 * \figWidth,\plotStartY) --  (\plotStartX + \x * \figWidth + \xTickOff * \figWidth + \x * \doaDist - 0.5 * \figWidth, \plotStartY - \majorTick);
        \node[anchor=north, rotate=0] at (\plotStartX + \x * \figWidth + \xTickOff * \figWidth + \x * \doaDist - 0.5 * \figWidth,\plotStartY) {\small \xLabel};
    }
}

\end{tikzpicture}
\vspace*{-20pt}
\caption{
Sample means of unprocessed (\picLegend{images/legend/unprocessed}) and adaptive, weakly guided extraction methods (\picLegend{images/legend/weak-seldnet.pdf},\,\picLegend{images/legend/weak-seldnet-ar-spatial.pdf},\,\picLegend{images/legend/weak-seldnet-ar-spatial-spectral.pdf}) evaluated on real-world recordings regarding speech quality (NISQA) and intelligibility (WER).
}
\label{fig:recordings}
\vspace*{-10pt}
\end{figure}

\section{Conclusion}
\label{sec:conclusion}
In this work, we proposed a novel speaker extraction approach in the Ambisonics domain for dynamic scenarios with moving speakers.
By introducing \textit{fixed} and \textit{adaptive} rotary steering, we could leverage the rotational invariance of Ambisonics for both tracking and enhancement, ensuring broad applicability across algorithmic implementations.
\textit{Autoregressively} incorporating the processed speech signal, we achieved substantial improvements particularly for closely spaced speakers. 
As a result, our method outperformed comparable non-autoregressive approaches and demonstrated robust enhancement in challenging real-world recordings with multiple speaker crossings and varying distance between speakers and array.

\bibliographystyle{IEEEtran}
\bibliography{strings,refs}

\begin{thebibliography}{10}
\providecommand{\url}[1]{#1}
\csname url@samestyle\endcsname
\providecommand{\newblock}{\relax}
\providecommand{\bibinfo}[2]{#2}
\providecommand{\BIBentrySTDinterwordspacing}{\spaceskip=0pt\relax}
\providecommand{\BIBentryALTinterwordstretchfactor}{4}
\providecommand{\BIBentryALTinterwordspacing}{\spaceskip=\fontdimen2\font plus
\BIBentryALTinterwordstretchfactor\fontdimen3\font minus
  \fontdimen4\font\relax}
\providecommand{\BIBforeignlanguage}[2]{{%
\expandafter\ifx\csname l@#1\endcsname\relax
\typeout{** WARNING: IEEEtran.bst: No hyphenation pattern has been}%
\typeout{** loaded for the language `#1'. Using the pattern for}%
\typeout{** the default language instead.}%
\else
\language=\csname l@#1\endcsname
\fi
#2}}
\providecommand{\BIBdecl}{\relax}
\BIBdecl

\bibitem{cherry53cocktail_party}
E.~C. Cherry, ``Some experiments on the recognition of speech, with one and two
  ears,'' \emph{J. Acoust. Soc. Am.}, vol.~25, 1953.

\bibitem{tesch24ssf_journal}
K.~Tesch and T.~Gerkmann, ``Multi-channel speech separation using spatially
  selective deep non-linear filters,'' \emph{IEEE/ACM TASLP}, vol.~32, 2024.

\bibitem{bohlender24sep_journal}
A.~Bohlender, A.~Spriet, W.~Tirry, and N.~Madhu, ``Spatially selective speaker
  separation using a {DNN} with a location dependent feature extraction,''
  \emph{IEEE/ACM TASLP}, vol.~32, 2024.

\bibitem{pandey12directional_speech_extraction}
A.~Pandey, S.~Lee, J.~Azcarreta, D.~Wong, and B.~Xu, ``All neural low-latency
  directional speech extraction,'' in \emph{Interspeech}, 2024.

\bibitem{gu24rezero}
R.~Gu and Y.~Luo, ``{ReZero}: Region-customizable sound extraction,''
  \emph{IEEE/ACM TASLP}, vol.~32, 2024.

\bibitem{zotter19ambisonics}
F.~Zotter and M.~Frank, \emph{Ambisonics: A Practical {3D} Audio Theory for
  Recording, Studio Production, Sound Reinforcement, and Virtual
  Reality}.\hskip 1em plus 0.5em minus 0.4em\relax Springer, 2019.

\bibitem{wang25rotary_steering}
S.~Wang, H.~Qiu, X.~Song, M.~Wang, and F.~Yao, ``Ambisonics neural speech
  extraction with directional feature and rotary steering,'' \emph{Applied
  Acoustics}, vol. 228, 2025.

\bibitem{yang23mcnet}
Y.~Yang, Q.~Changsheng, and X.~li, ``{McNet}: Fuse multiple cues for
  multichannel speech enhancement,'' in \emph{IEEE ICASSP}, 2023.

\bibitem{quan24spatialnet}
C.~Quan and X.~Li, ``{SpatialNet}: Extensively learning spatial information for
  multichannel joint speech separation, denoising and dereverberation,''
  \emph{IEEE/ACM TASLP}, vol.~32, 2024.

\bibitem{chen20libricss}
Z.~Chen, T.~Yoshioka, L.~Lu, T.~Zhou, Z.~Meng, Y.~Luo, J.~Wu, X.~Xiao, and
  J.~Li, ``Continuous speech separation: Dataset and analysis,'' in \emph{IEEE
  ICASSP}, 2020.

\bibitem{barker18chime5}
J.~Barker, S.~Watanabe, E.~Vincent, and J.~Trmal, ``The fifth {'CHiME'} speech
  separation and recognition challenge: Dataset, task and baselines,'' in
  \emph{Interspeech}, 2018.

\bibitem{kienegger25wg_ssf}
J.~Kienegger and T.~Gerkmann, ``Steering deep non-linear spatially selective
  filters for weakly guided extraction of moving speakers in dynamic
  scenarios,'' in \emph{Interspeech}, 2025.

\bibitem{kienegger25sg_ssf}
J.~Kienegger, A.~Mannanova, H.~Fang, and T.~Gerkmann, ``Self-steering deep
  non-linear spatially selective filters for efficient extraction of moving
  speakers under weak guidance,'' in \emph{IEEE WASPAA}, 2025.

\bibitem{luo19conv_tasnet}
Y.~Luo and N.~Mesgarani, ``{Conv-TasNet}: Surpassing ideal time–frequency
  magnitude masking for speech separation,'' \emph{IEEE/ACM TASLP}, vol.~27,
  2019.

\bibitem{chao24SEmamba}
R.~Chao, W.-H. Cheng, M.~L. Quatra, S.~M. Siniscalchi, C.-H.~H. Yang, S.-W. Fu,
  and Y.~Tsao, ``An investigation of incorporating {Mamba} for speech
  enhancement,'' in \emph{IEEE Spoken Language Tech. Workshop}, 2024.

\bibitem{andreev23iterative_autoregression}
P.~Andreev, N.~Babaev, A.~Saginbaev, I.~Shchekotov, and A.~Alanov, ``Iterative
  autoregression: A novel trick to improve your low-latency speech enhancement
  model,'' in \emph{Interspeech}, 2023.

\bibitem{pan24paris_autoregressive_separation}
Z.~Pan, G.~Wichern, F.~G. Germain, K.~Saijo, and J.~{Le Roux}, ``{PARIS}:
  Pseudo-autoregressive siamese training for online speech separation,'' in
  \emph{Interspeech}, 2024.

\bibitem{shen25arise}
P.~Shen, X.~Zhang, and Z.-Q. Wang, ``{ARiSE}: Auto-regressive multi-channel
  speech enhancement,'' in \emph{Interspeech}, 2025.

\bibitem{nachbar11ambix}
C.~Nachbar, F.~Zotter, E.~Deleflie, and A.~Sontacchi, ``{ambiX} - {A} suggested
  {Ambisonics} format,'' in \emph{Ambisonics Symposium}, 2011.

\bibitem{su94rotation_real_sh}
Z.~Su and P.~Coppens, ``Rotation of real spherical harmonics,''
  \emph{Foundations of Crystallography}, vol.~50, 1994.

\bibitem{rafaely08beam_steering_wignerD}
B.~Rafaely and M.~Kleider, ``Spherical microphone array beam steering using
  {Wigner-D} weighting,'' \emph{IEEE Signal Proc. Letters}, vol.~15, 2008.

\bibitem{battula25MIMO_localization}
S.~S. Battula, H.~Taherian, A.~Pandey, D.~Wong, B.~Xu, and D.~Wang, ``Robust
  frame-level speaker localization in reverberant and noisy environments by
  exploiting phase difference losses,'' in \emph{IEEE ICASSP}, 2025.

\bibitem{panayotov15librispeech}
V.~Panayotov, G.~Chen, D.~Povey, and S.~Khudanpur, ``Librispeech: An {ASR}
  corpus based on public domain audio books,'' in \emph{IEEE ICASSP}, 2015.

\bibitem{cosentino20librimix}
J.~Cosentino, M.~Pariente, S.~Cornell, A.~Deleforge, and E.~Vincent,
  ``Librimix: An open-source dataset for generalizable speech separation,''
  preprint arXiv:2005.11262, 2020.

\bibitem{allen79image_method}
J.~Allen and D.~Berkley, ``Image method for efficiently simulating small-room
  acoustics,'' \emph{J. Acoust. Soc. Am.}, vol.~65, 1979.

\bibitem{diaz18gpu_rir}
D.~Diaz-Guerra, A.~Miguel, and J.~R. Beltr{\'a}n, ``{gpuRIR}: A {Python}
  library for room impulse response simulation with {GPU} acceleration,''
  \emph{Multimedia Tools and Applications}, vol.~80, 2018.

\bibitem{rafaely19spherical_array_processing}
B.~Rafaely, \emph{Fundamentals of Spherical Array Processing}.\hskip 1em plus
  0.5em minus 0.4em\relax Springer, 2019.

\bibitem{diaz21srp_phat}
D.~Diaz-Guerra, A.~Miguel, and J.~R. Beltran, ``Robust sound source tracking
  using {SRP-PHAT} and {3D} convolutional neural networks,'' \emph{IEEE/ACM
  TASLP}, vol.~29, 2021.

\bibitem{fairbanks60rainbow_passage}
G.~Fairbanks, \emph{Voice and Articulation Drillbook}.\hskip 1em plus 0.5em
  minus 0.4em\relax Harper, 1960.

\bibitem{yasuda24causal_seldnet}
M.~Yasuda, S.~Saito, A.~Nakayama, and N.~Harada, ``{6DoF SELD}: Sound event
  localization and detection using microphones and motion tracking sensors on
  self-motioning human,'' in \emph{IEEE ICASSP}, 2024.

\bibitem{quan24online_spatialnet}
C.~Quan and X.~Li, ``Multichannel long-term streaming neural speech enhancement
  for static and moving speakers,'' \emph{IEEE Signal Proc. Letters}, vol.~31,
  2024.

\bibitem{adavanne19seldnet}
S.~Adavanne, A.~Politis, J.~Nikunen, and T.~Virtanen, ``Sound event
  localization and detection of overlapping sources using convolutional
  recurrent neural networks,'' \emph{IEEE J. Sel. Topics Signal Proc.},
  vol.~13, 2019.

\bibitem{rix01pesq}
A.~Rix, J.~Beerends, M.~Hollier, and A.~Hekstra, ``Perceptual evaluation of
  speech quality {(PESQ)}-a new method for speech quality assessment of
  telephone networks and codecs,'' in \emph{IEEE ICASSP}, 2001.

\bibitem{jensen16estoi}
J.~Jensen and C.~H. Taal, ``An algorithm for predicting the intelligibility of
  speech masked by modulated noise maskers,'' \emph{IEEE/ACM TASLP}, vol.~24,
  2016.

\bibitem{mittag21nisqa}
G.~Mittag, B.~Naderi, A.~Chehadi, and S.~M{\"o}ller, ``{NISQA}: A deep
  {CNN}-self-attention model for multidimensional speech quality prediction
  with crowdsourced datasets,'' in \emph{Interspeech}, 2021.

\bibitem{kuchaiev9nemo}
O.~Kuchaiev, J.~Li, H.~Nguyen, O.~Hrinchuk, R.~Leary, B.~Ginsburg, S.~Kriman,
  S.~Beliaev, V.~Lavrukhin, J.~Cook, P.~Castonguay, M.~Popova, J.~Huang, and
  J.~M. Cohen, ``{NeMo}: A toolkit for building {AI} applications using neural
  modules,'' 2019.

\end{thebibliography}

\end{document}